\documentstyle[12pt]{article}
\newfont{\bbd}{msbm10 scaled\magstep1}

\begin{document}
\thispagestyle{empty}

\def\ve#1{\mid #1\rangle}
\def\vc#1{\langle #1\mid}

\newcommand{\p}[1]{(\ref{#1})}
\newcommand{\be}{\begin{equation}}
\newcommand{\ee}{\end{equation}}
\newcommand{\sect}[1]{\setcounter{equation}{0}\section{#1}}

\newcommand{\vs}[1]{\rule[- #1 mm]{0mm}{#1 mm}}
\newcommand{\hs}[1]{\hspace{#1mm}}
\newcommand{\mb}[1]{\hs{5}\mbox{#1}\hs{5}}
\newcommand{\Db}{{\overline D}}
\newcommand{\bea}{\begin{eqnarray}}

\newcommand{\eea}{\end{eqnarray}}
\newcommand{\wt}[1]{\widetilde{#1}}
\newcommand{\und}[1]{\underline{#1}}
\newcommand{\ov}[1]{\overline{#1}}
\newcommand{\sm}[2]{\frac{\mbox{\footnotesize #1}\vs{-2}}
           {\vs{-2}\mbox{\footnotesize #2}}}
\newcommand{\prt}{\partial}
\newcommand{\eps}{\epsilon}

\newcommand{\R}{\mbox{\rule{0.2mm}{2.8mm}\hspace{-1.5mm} R}}
\newcommand{\Z}{Z\hspace{-2mm}Z}

\newcommand{\cd}{{\cal D}}
\newcommand{\cg}{{\cal G}}
\newcommand{\ck}{{\cal K}}
\newcommand{\cw}{{\cal W}}

\newcommand{\vj}{\vec{J}}
\newcommand{\vl}{\vec{\lambda}}
\newcommand{\vz}{\vec{\sigma}}
\newcommand{\vt}{\vec{\tau}}
\newcommand{\vw}{\vec{W}}
\newcommand{\poiss}{\stackrel{\otimes}{,}}

\def\l#1#2{\raisebox{.2ex}{$\displaystyle
  \mathop{#1}^{{\scriptstyle #2}\rightarrow}$}}
\def\r#1#2{\raisebox{.2ex}{$\displaystyle
 \mathop{#1}^{\leftarrow {\scriptstyle #2}}$}}



\renewcommand{\thefootnote}{\fnsymbol{footnote}}
\newpage
\setcounter{page}{0}
\pagestyle{empty}
\begin{flushright}
{June 1999}\\
{LPENSL--TH--14/99}\\
{solv-int/9907004}
\end{flushright}
\vfill

\begin{center}
{\LARGE {\bf N=2 local and N=4 nonlocal reductions}}\\[0.3cm]
{\LARGE {\bf of supersymmetric KP hierarchy}}\\[0.3cm]
{\LARGE {\bf in N=2 superspace}}\\[1cm]

{}~

{\large F. Delduc$^{a,1}$, L. Gallot$^{b,2}$ and A. Sorin$^{c,3}$}
{}~\\
\quad \\
{\em {~$~^{(a)}$ Laboratoire de Physique Th\'eorique ENSLAPP,}}\\
{\em ENS Lyon, 46 All\'ee d'Italie, 69364 Lyon, France}\\[10pt]
{\em {~$~^{(b)}$ Dipartimento di Fisica Teorica, Universit\`a di
Torino}}\\  {{\em and INFN, Sezione di Torino,}}\\
{\em Via Giuria 1, I-10125 Torino, Italy}\\[10pt]
{\em {~$~^{(c)}$ Bogoliubov Laboratory of Theoretical Physics, JINR,}}\\
{\em 141980 Dubna, Moscow Region, Russia}~\quad\\

\end{center}

\vfill

{}~

\centerline{{\bf Abstract}}
\noindent
A $N=4$ supersymmetric matrix KP hierarchy is proposed and a wide
class of its reductions which are characterized by a finite number
of fields are described. This class includes the one-dimensional
reduction of the two-dimensional $N=(2|2)$ superconformal Toda lattice
hierarchy possessing the $N=4$ supersymmetry -- the $N=4$ Toda chain
hierarchy -- which may be relevant in the construction of supersymmetric
matrix models. The Lax pair representations of the bosonic and fermionic
flows, corresponding local and nonlocal Hamiltonians, finite and infinite
discrete symmetries, the first two Hamiltonian structures and the
recursion operator connecting all evolution equations and the Hamiltonian
structures of the $N=4$ Toda chain hierarchy are constructed in explicit
form. Its secondary reduction to the $N=2$ supersymmetric ${\alpha}=-2$
KdV hierarchy is discussed.

{}~

{}~



{}~

{}~

\vfill
{\em E-Mail:\\
1) delduc@ens-lyon.fr\\
2) gallot@to.infn.it\\
3) sorin@thsun1.jinr.ru }
\newpage
\pagestyle{plain}
\renewcommand{\thefootnote}{\arabic{footnote}}
\setcounter{footnote}{0}

\section{Introduction}
Recently an infinite class of bosonic and fermionic solutions of the
symmetry equation of the two-dimensional $N=(1|1)$ superconformal Toda
lattice equation
was constructed in \cite{ls,ols}. These solutions generate bosonic and
fermionic flows of the $N=(1|1)$ supersymmetric Toda lattice hierarchy in
the same way as their bosonic counterparts -- solutions of the symmetry
equation of the Toda equation -- produce the flows of the bosonic Toda
lattice hierarchy. The symmetry equation is a complicated
nonlinear functional equation, and its general solution is not known.
In this respect one cannot exclude that beside the solutions found
in \cite{ls,ols} there are other solutions of the symmetry
equation which could be responsible for additional flows of the hierarchy. 
Moreover, most probably such solutions do exist. Indeed, if 
one remembers that the $N=(1|1)$ superconformal Toda lattice
equation actually admits an $N=(2|2)$ superconformal symmetry
\cite{ev} one can suppose that its one-dimensional reduction --
supersymmetric Toda chain equation -- possesses a global $N=4$ 
supersymmetry. In this respect it seems interesting to develop the
Lax-pair representation method for the supersymmetric Toda chain hierarchy
which is still unknown, aiming at the description of as many of its flows
as possible. This is the main goal of the present paper. 

Using a dressing formalism, we construct the Lax-pair description of the
supersymmetric Toda chain hierarchy and observe new series of both
fermionic and bosonic flows at our knowledge unknown before. We 
show that it represents a reduction of the supersymmetric KP hieararchy
in the $N=2$ superspace which actually possesses an $N=4$ supersymmetry.
We also derive its infinite-dimensional algebra of flows and demonstrate
that it indeed contains a global $N=4$ supersymmetry algebra including its
$gl(2)$ automorphisms. Due to this reason we call this hierarchy the $N=4$
supersymmetric Toda chain hierarchy. The Lax operators that we propose
possess a peculiarity which make them different from all other Lax
operators applied earlier to the description of integrable 
supersymmetric systems in the $N=2$ superspace (for recent papers, see 
\cite{lm}--\cite{dg3} and references therein): they are not invariant
with respect to the $U(1)$ automorphism of the $N=2$ supersymmetry. As a
byproduct we propose an infinite family of new reductions of the well
known  supersymmetric KP hierarchy in the $N=2$ superspace which lead to
$U(1)$ non-invariant Lax operators.  

The paper is organized as follows. In section 2 we develop tools for
the Lax-pair description of integrable systems with $N=4$ 
supersymmetry. Thus, in the framework of the dressing approach
we give the description of the supersymmetric KP hierarchy in the
$N=2$ superspace and demonstrate that
it actually is $N=4$ supersymmetric. In section 3
we construct a zero-curvature representation for the $N=(2|2)$
superconformal Toda lattice equation and use it to derive
the Lax operator generating the bosonic flows of the supersymmetric Toda
chain hierarchy. Then, in section 4 we use this Lax operator to
construct a consistent reduction of all other flows of the 
$N=4$ supersymmetric KP hierarchy preserving its algebra structure, 
and show that the reduced hierarchy -- supersymmetric Toda
chain -- possesses an $N=4$ supersymmetry.
In section 5 we discuss its finite and infinite discrete
symmetries, and use them to obtain new Lax operators. In section 6 we
construct its local and nonlocal Hamiltonians, the first two Hamiltonian
structures and the recursion operator.
In section 7 we obtain the $U(1)$--automorphism transformations for
all the flows and derive a new Lax operator. In section 8 we
establish a relationship between the $N=4$ Toda chain hierarchy and
$N=2$, ${\alpha}=-2$ KdV hierarchy. In section 9 we propose a
generalization of the $N=4$ KP and Toda chain hierarchies in the matrix
case. An appendix contains a realization of the algebra of flows of the
$N=4$ supersymmetric KP hierarchy.

\section{N=4 supersymmetric KP hierarchy}
In this section we discuss the hierarchy which is usually called the
$N=2$ supersymmetric KP hierarchy and focus on the
remarkable fact that it actually possesses an $N=4$
supersymmetry algebra. 

Our starting object is the $N=2$ supersymmetric dressing operator $W$
\begin{eqnarray}
W \equiv 1+\sum_{n=1}^\infty ~(w^{(0)}_n + w^{(+)}_nD_{+}+w^{(-)}_nD_{-}+
w^{(1)}_n D_{+}D_{-})~{\partial}^{-n},
\label{dreskp}
\end{eqnarray}
where all the functions $w_n\equiv w_n(Z)$ involved into $W$ are 
$N=2$ superfields in the superspace $Z\equiv (z,{\theta}^{+},{\theta}^{-})$, 
and $D_{\pm}$ are fermionic covariant derivatives which together with
the supersymmetry generators $Q_{\pm}$ form the 
algebra\footnote{Hereafter, we explicitly present only non-zero brackets.}
\begin{eqnarray}
\{ D_{\pm},D_{\pm}\} = +2{\partial}, \quad
\{ Q_{\pm},Q_{\pm}\} = -2{\partial}
\label{alg00}
\end{eqnarray}
with the standard superspace realization:
\begin{eqnarray}
D_{\pm}\equiv \frac{\partial}{\partial {\theta}^{\pm}} +
{\theta}^{\pm} {\partial}, \quad 
Q_{\pm}\equiv \frac{\partial}{\partial {\theta}^{\pm}} -
{\theta}^{\pm} {\partial}.
\label{algDQ0}
\end{eqnarray}

Our aim is to construct a maximal set of consistent Sato
equations for the dressing operator $W$ which are the flows
of the extended supersymmetric KP hierarchy in the $N=2$ superspace. 

Keeping this in mind, we first construct all linearly independent 
linear differential operators defined in the $N=2$ superspace which are 
then dressed by the operator $W$. As a result we obtain the operators
\begin{eqnarray}
&& \quad \quad \quad L_{l}^{\pm} \equiv W{D_{\pm}}^{l}W^{-1}, \quad
M_{l}^{\pm} \equiv W{Q_{\pm}}^{l}W^{-1}, \quad N_{l}^{\pm}\equiv 
W \frac{1}{2}[{\theta}^{\pm},~D_{\pm}]{\partial}^{l}W^{-1}, \nonumber\\
&& N_{l}\equiv - W
({\theta}^{+}D_{-} + {\theta}^{-}D_{+}){\partial}^{l}W^{-1},
\quad {\overline N}_{l}\equiv W\frac{1}{2}
({\theta}^{-}(D_{+}+Q_{+}) -
{\theta}^{+}(D_{-}+Q_{-})){\partial}^{l}W^{-1} ~~
\label{laxkp}
\end{eqnarray}
with the obvious properties: 
\begin{eqnarray}
L_{l}^{\pm} \equiv (L_{1}^{\pm})^{l}, \quad 
M_{l}^{\pm} \equiv (M_{1}^{\pm})^{l}, \quad  
L_{2l}^{+} \equiv L_{2l}^{-}\equiv (-1)^{l}M_{2l}^{+} 
\equiv (-1)^{l}M_{2l}^{-} 
\label{prop}
\end{eqnarray}
which will be useful in what follows.

Second, using the operators \p{laxkp} we construct the 
consistent Sato equations for $W$ 
\begin{eqnarray}
&& {\textstyle{\partial\over\partial t_l}}W= -(L_{2l}^{-})_-W, \quad
\quad ~U^{\pm}_{l}W= -(N_{l}^{\pm})_- W,\nonumber\\
&& U_{l}W=-(N_{l})_- W, \quad \quad \quad
{\overline U}_{l}W= -({\overline N}_{l})_-W, \nonumber\\
&&D^{\pm}_l W= -(L_{2l-1}^{\pm})_- W,\quad
~Q^{\pm}_lW= -(M_{2l-1}^{\pm})_- W,
\label{satokp}
\end{eqnarray}
where the subscript $-$ ( $+$ ) denote the purely pseudo-differential 
(differential) part of the operator. 
The bosonic (fermionic) evolution derivatives 
$\{{\textstyle{\partial\over\partial t_l}}, ~U^{\pm}_l, ~U_l, 
~{\overline U}_l\}$ (~$\{ D^{\pm}_l, ~Q^{\pm}_l\}$~) generating 
bosonic (fermionic) flows of the hierarchy under consideration
have the following length dimensions:
\begin{eqnarray}
[{\textstyle{\partial\over\partial t_l}}]=[U^{\pm}_l]=
[U_l]=[{\overline U}_l]=-l, \quad
[D^{\pm}_l]=[Q^{\pm}_l]=-l+\frac{1}{2}.
\label{dimtimes}
\end{eqnarray}

We would like to stress that the flows    
$\{{\textstyle{\partial\over\partial t_l}}, ~D^{\pm}_l\}$ form 
the hierarchy which is usually called the $N=2$ supersymmetric KP
hierarchy. Therefore, the flows $\{U^{\pm}_l, ~U_l, ~{\overline U}_l,
~Q^{\pm}_l\}$ when added to the $N=2$ KP hierarchy, produce an extended
hierarchy possessing a richer algebra structure.   

In order to understand deeper what is the extended hierarchy we have
in fact obtained, let us calculate its algebra of flows. One
can use a supersymmetric generalization \cite{stan,tak} of the Radul
map \cite{rad} which is a homomorphism between the algebra of
flows we are looking for and the algebra of the operators $L_{l}^{\pm}$,
$M_{l}^{\pm}$ , $N_{l}^{\pm}$, $N_{l}$ and ${\overline N}_{l}$ \p{laxkp}.
The resulting algebra is:
\begin{eqnarray}
\Bigl\{D^{\pm}_k\,,\,D^{\pm}_l\Bigr\}=
-2\;\frac{{\partial}}{{\partial t_{k+l-1}}}, \quad
\Bigl\{Q^{\pm}_k\,,\,Q^{\pm}_l\Bigr\}=
+2\;\frac{{\partial}}{{\partial t_{k+l-1}}}, 
\label{alg1}
\end{eqnarray}
\begin{eqnarray}
\Bigl[U_k\,,\,U^{\pm}_l\Bigr]= \pm {\overline U}_{k+l}, \quad
\Bigl[ {\overline U}_k\,,\,U^{\pm}_l\Bigr]=\mp U_{k+l}, \quad
\Bigl[U_k\,,\,{\overline U}_l\Bigr]=2 (U^{+}_{k+l}-U^{-}_{k+l}),
\label{algq}
\end{eqnarray}
\begin{eqnarray}
&&\Bigl[U^{\pm}_k\,,\,D^{\pm}_l\Bigr]=Q^{\pm}_{k+l}, \quad 
~\Bigl[U^{\pm}_k\,,\,Q^{\pm}_l\Bigr]=D^{\pm}_{k+l}, \nonumber\\
&&\Bigl[U_k\,,\,D^{\pm}_l\Bigr]=-Q^{\mp}_{k+l}, \quad
\Bigl[U_k\,,\,Q^{\pm}_l\Bigr]=-D^{\mp}_{k+l}, \nonumber\\
&&\Bigl[{\overline U}_k\,,\,D^{\pm}_l\Bigr]=\mp D^{\mp}_{k+l}, \quad
\Bigl[{\overline U}_k\,,\, Q^{\pm}_l\Bigr]=\mp Q^{\mp}_{k+l},
\label{algqqbar}
\end{eqnarray}
and its realization is given in the appendix.

A simple inspection of the superalgebra (\ref{alg1}--\ref{algqqbar})
shows that the flows ${\textstyle{\partial\over\partial t_1}}$,
$U^{\pm}_0$, $U_0$, ${\overline U}_0$, $D^{\pm}_1$ and $Q^{\pm}_1$
form the $N=4$ supersymmetry algebra including its $gl(2)$
automorphisms. This invariance algebra 
forms a finite--dimensional subalgebra of the 
superalgebra (\ref{alg1}--\ref{algqqbar}). Due to this reason the extended
hierarchy can be called the $N=4$ supersymmetric KP hierarchy.

To close this section let us only mention that a similar approach
with respect to the $N=1$ supersymmetric dressing operator was
developed in \cite{tak}, and as a result $N=2$ supersymmetric 
flows were derived. In \cite{tak} the corresponding supersymmetric
hierarchy was called the maximal SKP hierarchy, and it includes
both Manin-Radul \cite{mr} and Mulase-Rabin \cite{m,r} hierarchies.

\section{Hint for the N=4 KP reduction: zero-curvature representation 
of the $N=(2|2)$ superconformal Toda lattice}
In this section we derive the Lax operator generating the bosonic flows of
the supersymmetric Toda chain hierarchy using the zero-curvature
representation of the $N=(2|2)$ superconformal Toda lattice equation. 

One starts from the two-dimensional $N=(2|2)$ superconformal Toda lattice
equation
\begin{eqnarray}
D_-D_+\ln b_i=b_{i+1}-b_{i-1}
\label{zerocurveqs2}
\end{eqnarray}
written in terms of the bosonic $N=(1|1)$ superfield $b_i\equiv
b_i(z^+,\theta^+;z^-,\theta^-)$ defined on the lattice, 
$i\in \hbox{\bbd Z}$, and $D_{\pm}$ are the $N=1$ supersymmetric fermionic
derivatives in the right and left chiral subspaces 
$(z^{\pm},\theta^{\pm})$, 
\begin{eqnarray}
\{ D_{\pm},D_{\pm}\} = +2{\partial}_{\pm},\quad
D_{\pm}\equiv \frac{\partial}{\partial {\theta}^{\pm}} +
{\theta}^{\pm} {\partial}_{\pm}.
\label{alg0}
\end{eqnarray}
Eq. \p{zerocurveqs2} can be rewritten in the form of a system of two
equations \cite{ls}
\begin{eqnarray}
D_-f_i=b_i+b_{i+1}, \quad D_+\ln b_i=f_i-f_{i-1}
\label{zerocurveqs1}
\end{eqnarray}
which admits the zero-curvature representation
\begin{eqnarray}
\{D_+-A_{\theta^+}~,~D_--A_{\theta^-}\}=0
\label{zerocurv}
\end{eqnarray}
with the fermionic connections 
\begin{eqnarray}
(A_{\theta^+})_{ij}\equiv f_i{\delta}_{i,j}+{\delta}_{i,j-1}, \quad
(A_{\theta^-})_{ij} \equiv -b_i{\delta}_{i,j+1},
\label{fermconn}
\end{eqnarray}
where $f_i\equiv f_i(z^+,\theta^+;z^-,\theta^-)$ 
is a fermionic $N=(1|1)$ lattice superfield. One can 
define the bosonic connections $A_{z^{\pm}}$ by 
\begin{eqnarray}
{\partial}_+ +A_{z^+}\equiv (D_+-A_{\theta^+})^2, \quad
{\partial}_- +A_{z^-}\equiv (D_--A_{\theta^-})^2.
\label{boscon}
\end{eqnarray}
They explicitly read
\begin{eqnarray}
(A_{z^-})_{ij}\equiv D_-b_i{\delta}_{i,j+1} -b_ib_{i-1}{\delta}_{i,j+2},
\quad (A_{z^+})_{ij}\equiv -D_+f_i{\delta}_{i,j}+
(f_i-f_{i+1}){\delta}_{i,j-1} - {\delta}_{i,j-2}
\label{connect}
\end{eqnarray}
and due to \p{zerocurv} obviously satisfy the zero-curvature condition 
\begin{eqnarray}
[{\partial}_- + A_{z^-}~,~{\partial}_++A_{z^+}]=0 
\label{bozcon1}
\end{eqnarray}
which is a consistency condition for the following linear system:
\begin{eqnarray}
\quad ~~({\partial}_-+A_{z^-})\Psi=\lambda \Psi, \quad
\label{linsys1}
\end{eqnarray}
\begin{eqnarray}
({\partial}_++A_{z^+})\Psi=0,
\label{linsys2}
\end{eqnarray}
where $\Psi\equiv \Psi_i$ is the lattice wave function and $\lambda$ is a
spectral parameter. Taking into account the first relation of eqs.
\p{boscon}, equation \p{linsys2} can equivalently be rewritten in the
following form:
\begin{eqnarray}
~(D_+ -A_{\theta^+})\Psi=0.
\label{linsys3}
\end{eqnarray}
The linear system \p{linsys1}, \p{linsys3} is a key object in our
consideration. 

Now, let us consider a  consistent reduction to a one-dimensional
subspace
by setting
\begin{eqnarray}
{\partial}_-={\partial}_+\equiv \partial
\label{reduct}
\end{eqnarray}
which leads to the supersymmetric Toda chain. Then, in order to derive 
the Lax operator we are looking for, we follow a trick
proposed in \cite{bx} and express each lattice function entering the
spectral equation \p{linsys1} in terms of lattice functions defined at
the single lattice point $i$ using eqs. \p{zerocurveqs1}
and \p{linsys3}. Taking into account also the reduction condition
\p{reduct}, we obtain the following resulting spectral equation 
\begin{eqnarray}
\quad \quad (D_- + \frac{1}{D_{+}-f_i} b_i)^2\Psi_i=\lambda\Psi_i.
\label{lax1}
\end{eqnarray}
For each fixed value of $i$, it represents the spectral equation of
the differential supersymmetric Toda chain hierarchy, i.e. the
hierarchy of equations involving 
only the superfields $b_i,f_i$ at a single lattice point. The discrete
lattice shift (i.e., the system of eqs. \p{zerocurveqs1}) when added
to the differential hierarchy, generates the discrete supersymmetric Toda 
chain hierarchy \cite{ols}. Thus, the discrete hierarchy appears as a
collection of an infinite number of isomorphic differential hierarchies
\cite{bx}. 

It is well known that a spectral equation is just an equation for a Lax
operator. For a fixed value of $i$ one can omit the lattice index in
the spectral equation \p{lax1}, and it is obvious that the operator
\begin{eqnarray}
L=(D_- + \frac{1}{D_{+}-f} b)^2
\label{lax2}
\end{eqnarray}
entering eq. \p{lax1} is the Lax operator which is responsible for
the bosonic flows of the differential hierarchy. It can be simplified in
the new superfield basis $\{v_i,u_i\}$ defined as
\begin{eqnarray}
b_i\equiv u_iv_i, \quad f_i\equiv D_+\ln v_i
\label{zerocurveqs3}
\end{eqnarray}
in which the system \p{zerocurveqs1} takes the form of the $N=(1|1)$
supersymmetric generalization of the Darboux transformation \cite{ls}
\begin{eqnarray}
u_{i+1}=\frac{1}{v_{i}}, \quad
D_-D_+\ln v_i=u_{i+1}v_{i+1}+u_{i}v_{i},
\label{Darboux}
\end{eqnarray}
and the Lax operator \p{lax2} is
\begin{eqnarray}
L=(D_- + v D_{+}^{-1} u)^2.
\label{sollin}
\end{eqnarray}
This Lax operator will be used in the next section 
to construct a consistent reduction of all other flows of the
$N=4$ supersymmetric KP hierarchy.

\section{Reduction: bosonic and fermionic flows}
In this section we consider a reduction of the $N=4$ supersymmetric
KP hierarchy which is inspired by the Lax operator $L$ \p{sollin}
and preserves its algebra of flows (\ref{alg1}--\ref{algqqbar}).

Keeping in mind the results of the previous section and equation
\p{sollin}, let us introduce the following constraint on the operator 
$L_{1}^{-}$ \p{laxkp}
\begin{eqnarray} 
L_{1}^{-}= {\cal L} \equiv D_- + v D_{+}^{-1} u, 
\label{def1}
\end{eqnarray}
where ${\cal L}$ is the square root of the operator $L$ in eq. \p{sollin}.
The operator ${\cal L}$ possesses the following important 
properties\footnote{Let us recall the
operator conjugation rules: $D_{\pm}^{T}=-D_{\pm}$,
$(OP)^{T}=(-1)^{d_Od_P}P^{T}O^{T}$, where $O$ ($P$) is an arbitrary
operator with the Grassmann parity $d_O$ ($d_P$), and $d_O$=0 ($d_O=1$)
for bosonic (fermionic) operators $O$. All other rules can be
derived using these. Hereafter, we use the notation $(Of)$ for
an operator $O$ acting only on a function $f$ inside the brackets.}:
\begin{eqnarray}
({\cal L}^{2(l-1)})_-= \sum_{k=0}^{2l-3}
({\cal L}^{2l-3-k}v)D^{-1}_{+}(({\cal L}^{k})^{T}u), \quad l=0,1,2...,
\label{ar}
\end{eqnarray}
\begin{eqnarray}
({\cal L}^{2l-1})_-= \sum_{k=0}^{2(l-1)}({\cal L}^{2(l-1)-k}v)D^{-1}_{+}
(({\cal L}^{k})^{T}u)+ \sum_{k=0}^{2l-3}(-1)^{k}
({\cal L}^{2l-3-k}v)D^{-1}_{+}D_{-}(({\cal L}^{k})^{T}u)
\label{conj2}
\end{eqnarray}
which can be proved by induction in a similar way as analogous formulae in
the bosonic \cite{eor} and $N=1$ supersymmetric \cite{anp1} cases.
As one can see from eq. \p{ar}, even powers of ${\cal L}$ contain only the
operator $D_{+}$ (and not $D_{-}$) and are in some sense $N=1$ like.
Moreover, formula \p{ar} coincides with an analogous formula
in the $N=1$ supersymmetric case \cite{anp1}.

Substituting the expression \p{laxkp} for $L_{1}^{-}$ in
terms of the dressing operator $W$ \p{dreskp} into constraint \p{def1},
it becomes
\begin{eqnarray} 
WD_{-}W^{-1} = D_- + v D_{+}^{-1} u 
\label{def1eq}
\end{eqnarray}
and gives an equation for $W$ which can be solved iteratively 
with the following unique solution ${\cal W}$:  
\begin{eqnarray}
&& \quad \quad \quad \quad \quad 
{\cal W} \equiv W(w^{(-)}_{n}=0,~ w^{(1)}_{n}=0) \equiv
1+\sum_{n=1}^\infty ~(w^{(0)}_n + w^{(+)}_nD_{+})~{\partial}^{-n},\nonumber\\ 
&& \quad \quad \quad \quad \quad 
w^{(+)}_1\equiv -D_{-}^{-1}(uv), \quad
w^{(0)}_1\equiv -D_{-}^{-1}(vD_{+}u+uvD_{-}^{-1}(uv)), \nonumber\\
&&w^{(+)}_2\equiv -D_{-}^{-1}(vu~'+uvD_{+}D_{-}^{-1}(uv))+
(D_{-}^{-1}(uv))D_{-}^{-1}(vD_{+}u+uvD_{-}^{-1}(uv)), \quad \ldots
~~.~~~~~~
\label{def11}
\end{eqnarray}
Replacing $W$ by ${\cal W}$ in eqs. \p{laxkp} one can obtain
the reduced operators $M^{\pm}_l$,  $N^{\pm}_l$, $N_l$ and 
${\overline N}_l$ as well. As an example, we present a few terms of
the series in $D_{+}^{-1}$ in the case of $L^{+}_1$,
\begin{eqnarray}
L^{+}_{1}  \equiv  {\cal W}D_{+}{\cal W}^{-1} & = &
D_{+}+2w^{(+)}_1-(D_{+}w^{(+)}_1)D_{+}^{-1}
-((D_{+}w^{(0)}_1)-2w^{(+)}_2+w^{(+)}_1D_{+}w^{(+)}_1 \nonumber\\
&+& 2w^{(+)}_1 w^{(0)}_1)D_{+}^{-2} 
-((D_{+}(w^{(+)}_2-w^{(+)}_1w^{(0)}_1))-
(D_{+}w^{(+)}_1)^{2})D_{+}^{-3}+\ldots ~, ~~~~~
\label{def2}
\end{eqnarray}
where the functions $w^{(0)}_n$ and $w^{(+)}_n$ are defined in eqs.
\p{def11}. The following obvious relations:
\begin{eqnarray}
&&M^{-}_{1} = L^{-}_{1} - 2 {\theta}_{-} L^{-}_{2}, \quad  \quad
M^{+}_{1} = L^{+}_{1} - 2 {\cal W}{\theta}_{+}{\cal W}^{-1}
L^{+}_{2},\nonumber\\
&&N^{-}_{l} = {\theta}_{-}L^{-}_{2l+1} - \frac{1}{2} L^{-}_{2l},\quad  
N^{+}_{l} = {\cal W}{\theta}_{+}{\cal W}^{-1}L^{+}_{2l+1} - \frac{1}{2}
L^{+}_{2l}
\label{def3}
\end{eqnarray}
together with the relations \p{prop} are useful at calculations.

The most complicated task is to construct a consistent set of Sato
equations for the reduced ${\cal W}$ generalizing the unreduced equations 
\p{satokp} and preserving their algebra structure
(\ref{alg1}--\ref{algqqbar}).
A priori, it is completely unclear whether such equations exist at all.
Moreover, there are still no algorithmic methods to construct them.
Recently, a similar task was carried out in \cite{anp1} for some
reductions of the Manin-Radul
$N=1$ supersymmetric KP hierarchy \cite{mr}, and we use some of the ideas
developed there. We succeeded in this construction only
for the reduced ${\textstyle{\partial\over\partial t_l}}$, $U^{\pm}_l$,
$D^{\pm}_l$, and $Q^{\pm}_l$ flows. Nevertheless, at the end of this
section we propose a heuristic construction which allows the remaining 
$U_l$ and ${\overline U}_l$ flows of the reduced $N=4$ KP hierarchy 
to be restored as well. 

The resulting Sato equations have the following form:
\begin{eqnarray}
&& \quad \quad \quad  \quad \quad \quad 
{\textstyle{\partial\over\partial t_l}}{\cal W}=
-(L_{2l}^{-})_-{\cal W}, \nonumber\\
&&U^{+}_{l}{\cal W}=-(N_{l}^{+})_-{\cal W},\quad
~~U^{-}_{l}{\cal W}=-((N_{l}^{-})_- - 
{\widetilde N}^{-}_{2l+1}){\cal W}, \nonumber\\
&& D^{+}_l {\cal W}= -(L_{2l-1}^{+})_-{\cal W},\quad
D^{-}_l{\cal W}=-((L_{2l-1}^{-})_- - {\widetilde L}^{-}_{2l-1}){\cal W},
\nonumber\\ &&Q^{+}_l{\cal W}=-(M_{2l-1}^{+})_-{\cal W},\quad
Q^{-}_l{\cal W}=-((M_{2l-1}^{-})_- - {\widetilde M}^{-}_{2l-1}){\cal W},
\label{satoflow}
\end{eqnarray}
where new operators ${\widetilde L}^{-}_{2l-1}$, 
${\widetilde M}^{-}_{2l-1}$ and ${\widetilde N}^{-}_{2l-1}$
have been introduced,

{}~

\begin{eqnarray}
{\widetilde L}^{-}_{2l-1}\equiv 2\sum_{k=0}^{l-2}
({\cal L}^{2(l-k)-3}v)D^{-1}_{+}
(({\cal L}^{2k+1})^{T}u)+ \sum_{k=0}^{2l-3}(-1)^{k}
({\cal L}^{2l-3-k}v)D^{-1}_{+}D_{-}(({\cal L}^{k})^{T}u),
\label{conj3}
\end{eqnarray}
\begin{eqnarray}
{\widetilde M}^{-}_{2l-1}\equiv (-1)^{l-1}\sum_{k=0}^{2l-3}(-1)^{k}
({\cal L}^{2l-3-k}v)D^{-1}_{+}D_{-}(({\cal L}^{k})^{T}u),
\label{conj4}
\end{eqnarray}
\begin{eqnarray}
{\widetilde N}^{-}_{2l-1}\equiv -\sum_{k=0}^{l-2}
({\cal L}^{2(l-k-2)}v)D^{-1}_{+}(({\cal L}^{2k+1})^{T}u)
+{\theta}_{-}{\widetilde L}^{-}_{2l-1},
\label{conj5}
\end{eqnarray}
which are necessary for the consistency of the equations. 
The flows can easily be rewritten in the Lax-pair form,
\begin{eqnarray}
{\textstyle{\partial\over\partial t_l}}{\cal L} =
-[(L_{2l}^{-})_{-} ,{\cal L}]=[(L_{2l}^{-})_{+} ,{\cal L}],
\label{lax}
\end{eqnarray}
\begin{eqnarray}
&&\quad ~D^{+}_l{\cal L} = -\{ (L_{2l-1}^{+})_-,{\cal L}\}=\{
(L_{2l-1}^{+})_+,{\cal L}\}, \nonumber\\
&&\quad ~D^{-}_l{\cal L} = -\{ (L_{2l-1}^{-})_- - 
{\widetilde L}^{-}_{2l-1},{\cal L}\}= \{ (L^{-}_{2l-1})_{+}+
{\widetilde L}^{-}_{2l-1}, {\cal L}\} - 2{\cal L}^{2l},
\label{modfl2}
\end{eqnarray}
\begin{eqnarray}
&&Q^{+}_l{\cal L} =-\{ (M_{2l-1}^{+})_-,{\cal L}\}=
\{ (M_{2l-1}^{+})_+,{\cal L}\},\nonumber\\ &&Q^{-}_l{\cal L} =
-\{ (M_{2l-1}^{-})_- - {\widetilde M}^{-}_{2l-1},{\cal L}\}=
\{ (M_{2l-1}^{-})_{+}+{\widetilde M}^{-}_{2l-1}, {\cal L}\},
\label{etafl}
\end{eqnarray}
\begin{eqnarray}
&&~U^{+}_{l}{\cal L} =-[ (N_{l}^{+})_-,{\cal L}]=
[ (N_{l}^{+})_+,{\cal L}],\nonumber\\
&&~U^{-}_{l}{\cal L} = -[ (N_{l}^{-})_- - 
{\widetilde N}^{-}_{2l+1},{\cal L}]=
[-N_{l}^{-}+ (N_{l}^{-})_{+}+{\widetilde N}^{-}_{2l+1}, {\cal L}],
\label{qfl}
\end{eqnarray}
and lead to the following flow equations
for the superfields $v$ and $u$:
\begin{eqnarray}
&& {\textstyle{\partial\over\partial t_l}}v = ((L_{2l}^{-})_+v), \quad
{\textstyle{\partial\over\partial t_l}}u = -((L_{2l}^{-})^{T}_+u),
\nonumber\\ &&D^{+}_lv =((L_{2l-1}^{+})_+v),
\quad D^{+}_{l}u =-((L_{2l-1}^{+})^{T}_+u), \nonumber\\
&&D^{-}_{l}v =(((L_{2l-1}^{-})_{+}+
{\widetilde L}^{-}_{2l-1} - 2L_{2l-1}^{-})v), \nonumber\\
&&D^{-}_{l}u = -(((L_{2l-1}^{-})^{T}_{+}+({\widetilde L}^{-}_{2l-1} -
2L_{2l-1}^{-})^{T})u),
\nonumber\\ &&Q^{+}_{l}v =((M_{2l-1}^{+})_+v),\quad 
Q^{+}_{l}u =-((M_{2l-1}^{+})^{T}_+u),\nonumber\\
&&Q^{-}_{l}v =(((M_{2l-1}^{-})_{+}+{\widetilde M}^{-}_{2l-1})v),
\nonumber\\
&&Q^{-}_{l}u =-(((M_{2l-1}^{-})^{T}_{+}+({\widetilde
M}^{-}_{2l-1})^{T})u),
\nonumber\\ &&U^{+}_{l}v =((N_{l}^{+})_+v),\quad 
U^{+}_{l}u =-((N_{l}^{+})^{T}_+u),\nonumber\\
&&U^{-}_{l}v =(((N_{l}^{-})_{+}+{\widetilde N}^{-}_{2l+1} - 2N_{l}^{-})v), 
\nonumber\\ &&U^{-}_{l}u = -(((N_{l}^{-})^{T}_{+}+
({\widetilde N}^{-}_{2l+1} - 2N_{l}^{-})^{T})u),
\label{modflv}
\end{eqnarray}
where the equations (\ref{ar}--\ref{conj2}) have been used.

Using eqs. \p{modflv} for the bosonic and fermionic
flows, we present for illustration the first few\footnote{We have
rescaled some evolution derivatives to simplify the presentation
of some formulae.},
\begin{eqnarray}
&&\quad \quad \quad \quad  {\textstyle{\partial\over\partial t_0}}
\left(\begin{array}{cc} v\\ u \end{array}\right) =
\left(\begin{array}{cc} +v\\ -u \end{array}\right), \quad
{\textstyle{\partial\over\partial t_1}}
\left(\begin{array}{cc} v\\ u \end{array}\right) =
{\partial}\left(\begin{array}{cc} v\\ u \end{array}\right),
\nonumber\\ &&{\textstyle{\partial\over\partial t_2}} v =
+v~'' +  2uv(D_{+}D_{-}v)-(D_{+}D_{-}v^2u)-v^2(D_{+}D_{-}u) +2v(uv)^2,
\nonumber\\
&&{\textstyle{\partial\over\partial t_2}} u =
-u~'' +  2uv(D_{+}D_{-}u)-(D_{+}D_{-}u^2v)-u^2(D_{+}D_{-}v) -2u(uv)^2,
\label{eqs}
\end{eqnarray}
\begin{eqnarray}
&&\quad \quad \quad D^{\pm}_1 v= -D_{\pm}v\pm 2vD^{-1}_{\mp}(uv), 
\quad D^{\pm}_1 u= -D_{\pm}u\mp 2uD^{-1}_{\mp}(uv), \nonumber\\
&& D^{\pm}_2 v =
-D_{\pm}v~'\pm 2v~'D^{-1}_{\mp}(uv) \pm (D_{\pm}v)D^{-1}_{\mp}D_{\pm}(uv)
\pm vD^{-1}_{\mp}[uv~' +(D_{\pm}v)D_{\pm}u],\nonumber\\
&&D^{\pm}_2 u =+D_{\pm}u~'\pm 2u~'D^{-1}_{\mp}(uv) \pm 
(D_{\pm}u)D^{-1}_{\mp}D_{\pm}(uv)
\pm uD^{-1}_{\mp}[vu~' + (D_{\pm}u)D_{\pm}v],
\label{ff2-}
\end{eqnarray}
\begin{eqnarray}
\quad \quad Q^{\pm}_1 \left(\begin{array}{cc} v\\ u \end{array}\right) =
Q_{\pm}\left(\begin{array}{cc} v\\ u \end{array}\right),
\label{supersflows}
\end{eqnarray}
\begin{eqnarray}
\quad \quad U^{\pm}_0 v=\frac{1}{2}v-{\theta}^{\pm}
(D_{\pm}v\mp 2vD^{-1}_{\mp}(uv)), \quad U^{\pm}_0 u=\frac{1}{2}u -
{\theta}^{\pm} (D_{\pm}u\pm 2uD^{-1}_{\mp}(uv)).
\label{q1}
\end{eqnarray}
The flows (\ref{eqs}--\ref{supersflows}) reproduce the one-dimensional
reduction of the flows of the two-dimensional $N=(1|1)$ superconformal
Toda lattice hierarchy derived in \cite{ls,ols}. 

As it was already announced at the beginning of this section, now we
would like to briefly discuss the construction of the remaining two series
of the $N=4$ KP flows, i.e. $U_k$ and ${\overline U}_k$, for the reduced
hierarchy. This construction is based on the very simple and obvious
observation that the first two 
commutation relations of the subalgebra \p{algq} of the algebra
(\ref{alg1}--\ref{algqqbar}) can be used to construct
the flows $U_k$ and ${\overline U}_k$ for $k=1,2,\ldots$ in terms of
the flows $U^{+}_k$ already known and  presented in eqs. \p{modflv} 
as well as the flows $U_0$ and ${\overline U}_0$  
\begin{eqnarray}
&&U_0 v= -{\theta}^{+}(D_{-}v+ 2vD^{-1}_{+}(uv))
-{\theta}^{-}(D_{+}v- 2vD^{-1}_{-}(uv)), \nonumber\\
&&U_0 u=-{\theta}^{+}(D_{-}u- 2uD^{-1}_{+}(uv))
-{\theta}^{-}(D_{+}u+2uD^{-1}_{-}(uv)),
\label{qq1}
\end{eqnarray}
\begin{eqnarray}
\quad \quad \quad
{\overline U}_0 \left(\begin{array}{cc} v\\ u \end{array}\right) =
\frac{1}{2}~(~{\theta}^{-}(D_{+}+Q_{+})-{\theta}^{+}(D_{-}+Q_{-})~)
\left(\begin{array}{cc} v\\ u \end{array}\right)
\label{qqq1}
\end{eqnarray}
which were derived by brute force in order to satisfy the algebra
(\ref{alg1}--\ref{algqqbar}). Therefore, due to existence of the flows 
$U_0$ and ${\overline U}_0$ (\ref{qq1}--\ref{qqq1}), the flows $U_k$ and
${\overline U}_k$ for $k=1,2,\ldots$ can indeed be obtained,
and they explicitly read:
\begin{eqnarray}
\quad \quad U_{k}=-\Bigl[ {\overline U}_0\,,\,U^{+}_k\Bigr], \quad 
{\overline U}_{k}= \Bigl[U_0\,,\,U^{+}_k\Bigr].
\label{genrel}
\end{eqnarray}

We would like to close this section with a few remarks.

First, a simple inspection of eqs. \p{modflv} shows that  
the reduced flows differ from the $N=4$ KP flows in that almost all flows
except the bosonic flows ${\textstyle{\partial\over\partial t_l}}$
are nonlocal (for an example, see eqs.
\p{ff2-}, (\ref{q1}--\ref{qq1}) ). Nonlocal fermionic flows of
supersymmetric KdV, GNLS and Toda type systems were discussed also earlier
in \cite{ker,dmat,ra,bd,bs,anp1,ols} (see also the quite recent paper
\cite{mm}).   

Second, the flows $\{{\textstyle{\partial\over\partial t_1}},
~U^{\pm}_0,~U_0,~{\overline U}_0,~D^{\pm}_1,~Q^{\pm}_1\}$ 
forming the $N=4$ supersymmetry algebra are non-locally and
non-linearly realized in terms of the initial superfields $v$ and $u$.
However, there exists another superfield basis 
$\{{\widehat v},{\widehat u}\}$, defined as
\begin{eqnarray}
\{v,\quad u\} \quad \Longrightarrow \quad \{~{\widehat v}\equiv
v \exp{[-D_{+}^{-1}D_{-}^{-1}(uv)]}, \quad {\widehat u}\equiv
u\exp{[+D_{+}^{-1}D_{-}^{-1}(uv)]}~\},
\label{N4lintransf}
\end{eqnarray}
which localizes and linearizes the $N=4$ supersymmetry realization which
becomes now
\begin{eqnarray}
&&\quad \quad \quad  {\textstyle{\partial\over\partial t_1}}={\partial},
\quad D^{\pm}_1 =-D_{\pm}, \quad Q^{\pm}_1 =Q_{\pm}, \quad
U^{\pm}_0 = \frac{1}{2}~[D_{\pm},~{\theta}^{\pm} ], \nonumber\\
&&U_0 =-~(~{\theta}^{+}D_{-}+{\theta}^{-}D_{+}~), \quad
{\overline U}_0 =
\frac{1}{2}~(~{\theta}^{-}(D_{+}+Q_{+})-{\theta}^{+}(D_{-}+Q_{-})~).
\label{N4linflows}
\end{eqnarray}
However, in this basis the even flows  
${\textstyle{\partial\over\partial t_l}}$ for $l \geq 2$ are nonlocal.

Third, the flows $Q^{\pm}_l$ for $l \geq 2$ as well as the flows
$U^{\pm}_l$, $U_l$ and ${\overline U}_l$ in eqs. \p{modflv} and \p{genrel} 
are obtained at our knowledge for the first
time together with the Lax-pair representation of the whole hierarchy. 

Let us finally stress that, contrary to all known Lax operators
used before in the literature for the description of integrable 
supersymmetric systems in the $N=2$ superspace, the Lax operators
proposed in this section are not invariant with respect to the
$U(1)$--automorphism of the $N=2$ supersymmetry. We will return in section
7 to the discussion of consequences of this important difference.      

\section{Discrete symmetries, Darboux-B\"acklund transformations
and solutions}
In this section we discuss finite and infinite discrete
symmetries of the reduced hierarchy, and use them to construct
its solutions and new Lax operators. 

Direct verification shows that the flows (\ref{eqs}--\ref{qqq1})
admit the four involutions:
\begin{eqnarray}
&&\quad \quad \quad \quad (v,u)^{*}= i(u,v), \quad
(z,{{\theta}^{\pm}})^{*}=(z,{\theta}^{\pm}), \nonumber\\
&&(t_p,U^{\pm}_p,U_p,{\overline U}_p, D^{\pm}_p,
Q^{\pm}_p)^{*}=(-1)^{p-1}(t_p,-U^{\pm}_p,-U_p,-{\overline U}_p,
D^{\pm}_p,Q^{\pm}_p), ~ ~ ~ ~ ~
\label{conj*}
\end{eqnarray}
\begin{eqnarray}
&&\quad \quad \quad \quad (v,u)^{\dagger}= (u,v), \quad
(z,{{\theta}^{\pm}})^{\dagger}=(z,{\theta}^{\mp}), \nonumber\\
&&(t_p,U^{\pm}_p,U_p,{\overline U}_p,
D^{\pm}_p,Q^{\pm}_p)^{\dagger}=(-1)^{p-1}
(t_p,-U^{\mp}_p,-U_p,{\overline U}_p,
D^{\mp}_p,Q^{\mp}_p), ~ ~ ~ ~ ~
\label{conjdagger}
\end{eqnarray}
\begin{eqnarray}
&&\quad \quad \quad \quad (v,u)^{\star}= (v,u), \quad
(z,{{\theta}^{\pm}})^{\star}=(z,\pm {\theta}^{\mp}),\nonumber\\
&&(t_p,U^{\pm}_p,U_p,{\overline U}_p,D^{\pm}_p,
Q^{\pm}_p)^{\star}=(t_p,U^{\mp}_p,
-U_p,{\overline U}_p,\pm D^{\mp}_p,\pm Q^{\mp}_p),
\label{conjstar}
\end{eqnarray}
\begin{eqnarray}
&& \quad \quad \quad \quad
v^{\bullet}= v\exp{[-({\theta}^{+}(D_{-}^{-1}-Q_{-}^{-1})-
{\theta}^{-}(D_{+}^{-1}-Q_{+}^{-1}))(uv)]}, \nonumber\\
&& \quad \quad \quad \quad
u^{\bullet}= u\exp{[+({\theta}^{+}(D_{-}^{-1}-Q_{-}^{-1})-
{\theta}^{-}(D_{+}^{-1}-Q_{+}^{-1}))(uv)]}, \nonumber\\
&&(z,{{\theta}^{\pm}})^{\bullet}=- (z,{\theta}^{\pm}), \quad
(t_p,U^{\pm}_p,U_p,{\overline U}_p,D^{\pm}_p,
Q^{\pm}_p)^{\bullet}=(-1)^{p}(t_p,U^{\pm}_p,U_p,
{\overline U}_p,-Q^{\pm}_p,-D^{\pm}_p) ~~~~
\label{conbullet}
\end{eqnarray}
which are consistent with their algebra (\ref{alg00}),
(\ref{alg1}--\ref{algqqbar}). 

It is a simple exercise to check that all the flows (\ref{lax}--\ref{qfl})
(or \p{modflv} ) also possess the involution \p{conj*}, using the
following involution property of the dressing operator ${\cal W}$: 
\begin{eqnarray}
{\cal W}^{*}= ({\cal W}^{-1})^{T}
\label{Wconj*}
\end{eqnarray}
resulting from eq. \p{def1eq} and its consequences 
\begin{eqnarray}
&&({L^{\pm}_l})^{*}= (-1)^{\frac{l(l+1)}{2}}(L^{\pm}_l)^{T}, \quad 
({M^{\pm}_l})^{*}= (-1)^{\frac{l(l+1)}{2}}(M^{\pm}_l)^{T}, \quad 
({N^{\pm}_l})^{*}= (-1)^{\frac{l(l+1)}{2}}(N^{\pm}_l)^{T}, \nonumber\\
&&({{\widetilde L}^{-}_{2l-1}})^{*}= (-1)^{l}
({\widetilde L}^{-}_{2l-1})^{T}, \quad 
({{\widetilde M}^{-}_{2l-1}})^{*}=
(-1)^{l}({\widetilde M}^{-}_{2l-1})^{T}, \quad
({{\widetilde N}^{-}_{2l-1}})^{*}=
(-1)^{l}({\widetilde N}^{-}_{2l-1})^{T}
\label{opconj*}
\end{eqnarray}
for the operators entering eqs. (\ref{lax}--\ref{qfl}). As regards
the involutions (\ref{conjdagger}--\ref{conbullet}), we do not
have a direct proof that they are symmetries of all flows due to the 
complicated transformation properties of the dressing operator.
Fortunately, a simple proof can be given using the recurrence
relations \p{recrel}, to be derived later. We return to this question in
section 6 (see the paragraph after eq. \p{dens}).   

Beside involutions (\ref{conj*}--\ref{conbullet}), flows
(\ref{eqs}--\ref{qqq1}) possess the infinite-dimensional group of discrete
Darboux transformations \p{Darboux} \cite{ls,ols}
\begin{eqnarray}
&&(v,~u)^{\ddagger}=(~v(D_-D_+\ln v-uv),~ \frac{1}{v}~), \quad
(z,{{\theta}^{\pm}})^{\ddagger}=(z,{\theta}^{\pm}),\nonumber\\
&&(t_p,U^{\pm}_p,U_p,{\overline U}_p,D^{\pm}_p,
Q^{\pm}_p)^{\ddagger}=(t_p,-U^{\pm}_p,
-U_p,{\overline U}_p,-D^{\pm}_p,Q^{\pm}_p)
\label{discrsymm}
\end{eqnarray}
which may be written on the Lax operator ${\cal L}$ \p{def1} 
as\footnote{For the reduced Manin-Radul $N=1$
supersymmetric KP hierarchy the Darboux-B\"acklund transformations were
discussed in \cite{anp1} (see also references there).}:
\begin{eqnarray}
{\cal L}^{(\ddagger)} = - {\cal T}{\cal L}{\cal T}^{-1},
\quad {\cal T}\equiv v D_{+} v^{-1}.
\label{DBtransform1}
\end{eqnarray}

Applying involutions (\ref{conj*}--\ref{conbullet})
and the discrete group \p{discrsymm} to the Lax operator
${\cal L}$ \p{def1} one can derive other consistent Lax operators
\begin{eqnarray}
&&{\cal L}^{*}= D_- - u D_{+}^{-1} v \equiv -{\cal L}^{T}, \quad
\quad {\cal L}^{\dagger} = D_+ + u D_{-}^{-1} v, \nonumber\\
&&{\cal L}^{\star}= - D_+ + v D_{-}^{-1} u \equiv ({\cal L}^{\dagger})^{T},
\quad {\cal L}^{\bullet}= - Q_- + v^{\bullet} Q_{+}^{-1} u^{\bullet},
\label{conjlax}
\end{eqnarray}
\begin{eqnarray}
{\cal L}^{((j+1)\ddagger)} = D_- +
v^{((j+1)\ddagger)} D_{+}^{-1} u^{((j+1)\ddagger)}\equiv -
{\cal T}^{(j\ddagger)}{\cal L}^{(j\ddagger)}{{\cal T}^{(j\ddagger)}}^{-1},
\quad {\cal T}^{(j\ddagger)}\equiv v^{(j\ddagger)} D_{+}
{v^{(j\ddagger)}}^{-1}
\label{DBtransform}
\end{eqnarray}
generating isomorphic flows. 
${\cal L}^{(j\ddagger)}$ is obtained from ${\cal L}$ by applying  
$j$ times the discrete transformation \p{DBtransform1}, e.g.
${\cal L}^{(3\ddagger)} \equiv (({\cal L}^{\ddagger})^{\ddagger})
^{\ddagger}$, ${\cal L}^{(0\ddagger)} \equiv {\cal L}$. 

One can construct an infinite class of solutions of the reduced
hierarchy under consideration generalizing the results obtained in
\cite{ols} for the bosonic and fermionic flows $\frac{{\partial}}{\partial
t_k}$ and $D^{\pm}_l$. These results are based
on the discrete Darboux symmetry \p{discrsymm}, and we present their
generalization with brief comments referring to \cite{ols} for details.

The simplest solution of the hierarchy is:
\begin{eqnarray}
u = 0,
\label{boncond}
\end{eqnarray}
then the bosonic and fermionic flows for the remaining superfield $v\equiv
-{\tau}_0$ are linear and have the following form:
\begin{eqnarray}
{\textstyle{\partial\over\partial t_k}}{\tau}_0={\partial}^{k}{\tau}_0,
\quad D^{\pm}_k{\tau}_0=-D_{\pm} {\partial}^{k-1}{\tau}_0, \quad
Q^{\pm}_k{\tau}_0=Q_{\pm}{\partial}^{k-1}{\tau}_0, 
\label{eqv}
\end{eqnarray}
\begin{eqnarray}
&&U^{\pm}_k{\tau}_0=
\frac{1}{2}~[D_{\pm},~{\theta}^{\pm}]{\partial}^{k}{\tau}_0, 
\quad U_k{\tau}_0=
-~(~{\theta}^{+}D_{-}+{\theta}^{-}D_{+}~){\partial}^{k}{\tau}_0,
\nonumber\\ && \quad \quad {\overline U}_k{\tau}_0=
\frac{1}{2}~(~{\theta}^{-}(D_{+}+Q_{+})-{\theta}^{+}(D_{-}+Q_{-})~)
{\partial}^{k}{\tau}_0. 
\label{eqv1}
\end{eqnarray}
To derive these equations it is only necessary to take into account the
length dimensions \p{dimtimes} of the evolution derivatives, their algebra
(\ref{alg1}--\ref{algqqbar}) and the invariance of all flows
(\ref{lax}--\ref{qfl}) with respect to the following $U(1)$
transformations
\begin{eqnarray}
(~v,\quad u~) \quad \Longrightarrow \quad 
(~\exp{(+i\beta)}~v,\quad \exp{(-i\beta)}~u~)
\label{u1}
\end{eqnarray}
(consequently, only linear equations for $v$ are admissible at $u=0$)
which is obvious due to the invariance of the reduction constraint 
\p{def1eq}. $\beta$ is an arbitrary parameter. 

Here, in order to simplify formulae, we restrict the analysis of the
whole hierarchy to the case when only the flows 
$\{\frac{{\partial}}{\partial t_k}, D^{\pm}_k, Q^{\pm}_k\}$
are considered. Then, using the realization \p{covder} of the appendix 
the solution of eqs. \p{eqv} is
\begin{eqnarray}
\tau_{0} &=& \int\!\! d\lambda\; d\eta_{+}\; d\eta_{-}\;
\varphi(\lambda,\eta_+,\eta_-)\; \nonumber\\
&\times & \exp\Bigl\{ x{\lambda} {-}\!\sum_{\alpha=\pm}
{\eta}_{\alpha} {\theta}^{\alpha} {+}\! \sum^{\infty}_{k=1} \Bigl [t_k
{+}\!\sum_{\alpha=\pm}\Bigl (
\eta_{\alpha}({\theta}^{\alpha}_k-{\rho}^{\alpha}_k){\lambda}^{-1}+
\theta^{\alpha}({\theta}^{\alpha}_k+{\rho}^{\alpha}_k)-
{\theta}^{\alpha}_k \sum^{\infty}_{n=1} {\rho}^{\alpha}_n
{\lambda}^{n-1}\Bigr ) \Bigr]{\lambda}^{k}\Bigr \} ~~~~
\label{gensol}
\end{eqnarray}
where $\varphi$ is an arbitrary function of the
bosonic ($\lambda$) and fermionic ($\eta_{\pm}$)
spectral parameters with length dimensions
\begin{eqnarray}
[\lambda]= -1, \qquad
[\eta_{\pm}] = - \frac{1}{2}.
\label{dim1}
\end{eqnarray}
Applying the discrete group \p{discrsymm} to the solution constructed
$\{u=0,~v=-{\tau}_0\}$, an infinite
class of new solutions of the hierarchy is generated through an
obvious iterative procedure \cite{ols}
\begin{eqnarray}
&&v^{((2j+1)\ddagger)}\ =\ +(-1)^{j}\frac{\tau_{2j}}{\tau_{2j+1}}, \qquad\
\quad v^{(2(j+1)\ddagger)}\ =\
(-1)^{j}\frac{\tau_{2(j+1)}}{\tau_{2j+1}},
\nonumber\\[8pt]
&&u^{((2j+1)\ddagger)}\ =\ -(-1)^{j}\frac{\tau_{2j-1}}{\tau_{2j}}, \qquad
\quad u^{(2(j+1)\ddagger)}\ =\ (-1)^{j}\frac{\tau_{2j+1}}{\tau_{2j}},
\quad j=0,1,2, \ldots ,
\label{todasol}
\end{eqnarray}
where the $\tau_j$ are\footnote{ The superdeterminant is defined as
${\quad \rm sdet} \left(\begin{array}{cc} A & B \\ C & D
\end{array}\right)\ \equiv\
\det (A-BD^{-1}C ) (\det D)^{-1}$.}
\begin{eqnarray}
&& \tau_{2j}\quad\ =\ {\rm sdet} \biggl(\begin{array}{cc}
{\partial}^{p+q}\tau_0
& {\partial}^{p+m}D_-\tau_0 \\
{\partial}^{k+q}D_+\tau_0
& {\partial}^{k+m}D_+D_-\tau_0
\end{array}\biggr)^{0\leq p,q \leq j}_{0\leq k,m \leq j-1},
\nonumber\\[10pt]
&&\tau_{2j+1}\ =\ {\rm sdet} \biggl(\begin{array}{cc}
{\partial}^{p+q}\tau_0
& {\partial}^{p+m}D_-\tau_0 \\
{\partial}^{k+q}D_+\tau_0
& {\partial}^{k+m}D_+D_-\tau_0
\end{array}\biggr)^{0\leq p,q \leq j}_{0\leq k,m \leq j}.
\label{supdet}
\end{eqnarray}

\section{Hamiltonian structure}
In this section we construct local and nonlocal Hamiltonians,
the first two Hamiltonian structures and the recursion operator
of the reduced hierarchy.

Let us first present our notations for the $N=2$ superspace measure and
delta function
\begin{eqnarray}
dZ \equiv dz d \theta^{+} d \theta^{-}, \quad
{\delta}^{N=2}(Z) \equiv \theta^{+} \theta^{-} {\delta}(z),
\label{notat0}
\end{eqnarray}
respectively, as well as the realization of the inverse derivatives
\begin{eqnarray}
&&D_{\pm}^{-1}\equiv D_{\pm}{\partial}_{z}^{-1}, \quad
Q_{\pm}^{-1}\equiv -Q_{\pm}{\partial}_{z}^{-1}, \quad
{\partial}_{z}^{-1} \equiv \frac{1}{2}
\int_{-\infty}^{+\infty}dx{\epsilon}(z-x), \nonumber\\
&& \quad \quad \quad \quad
{\epsilon}(z-x)=-{\epsilon}(x-z)\equiv 1, \quad if \quad z>x
\label{derrealiz}
\end{eqnarray}
which we use in what follows. We also use the correspondence:
\begin{eqnarray}
{\textstyle{\partial\over\partial {\tau}^{a}_l}}
\equiv \{ {\textstyle{\partial\over\partial t_l}},U^{\mp}_l,U_l,
{\overline U}_l, D^{\mp}_l, Q^{\mp}_l\}
\quad \Leftrightarrow \quad
{\cal H}^{a}_l \equiv \{{\cal H}^{t}_l,
{\cal H}^{U^{\mp}}_l,{\cal H}^{U}_l, {\cal H}^{{\overline U}}_l,
{\cal H}^{D^{\mp}}_l, {\cal H}^{Q^{\mp}}_l\} 
\label{notat}
\end{eqnarray}
between the evolution derivatives
${\textstyle{\partial\over\partial {\tau}^{a}_l}}$ and Hamiltonian
densities ${\cal H}^{a}_l$. Consequently, the last ones have the
length dimensions:
\begin{eqnarray}
[{\cal H}^{t}_l]=
[{\cal H}^{U^{\mp}}_l]=[{\cal H}^{U}_l]=[{\cal H}^{{\overline U}}_l]
=-l, \quad [{\cal H}^{{D}^{\mp}}_l]=
[{\cal H}^{{Q}^{\mp}}_l]=-l+\frac{1}{2}.
\label{hams0}
\end{eqnarray}
The length dimensions of the Hamiltonian $H^{a}_l$,
\begin{eqnarray}
{H}^{a}_l \equiv \int d Z {\cal H}^{a}_l,
\label{hams00}
\end{eqnarray}
and its density ${\cal H}^{a}_l$ coincide because the $N=2$ superspace
measure has zero length dimension, $[dZ]=0$.

Let us next discuss a proper definition of the residue for the class of 
pseudo-differential operators generated by powers of the Lax operators
proposed in section 4. As we already mentioned, even powers of our Lax
operator ${\cal L}$ \p{def1} look like pure $N=1$ supersymmetric
operators, since they do not contain the operator $D_-$ (see paragraph
after eq. \p{conj2}). Due to this reason it seems reasonable to suppose
that the residue have to coincide with the residue usually used for an
$N=1$ supersymmetric pseudo-differential operator, i.e. with the
coefficient of the operator $D_{+}^{-1}$. Remembering that
this coefficient, when integrated over the $N=1$ supersymmetric
measure $dZ_+\equiv dz d{\theta}_{+}$, has to be actually $N=2$
supersymmetric as the hierarchy we started with, it should
necessarily admit a representation in the form of the operator
$D_{-}$ acting on a functional $F(z,\theta_+,\theta_-)$ of the superfields
$v$ and $u$ of the hierarchy. Indeed, in this case the equality  
$\int dZ_{+}(D_{-}F)(z,\theta_+,0)\equiv 
\int dZ F(z,\theta_+,\theta_-)$ is obviously valid just due to
the standard definition of the $N=2$ superspace integral
and, consequently, as a result we obtain an $N=2$ supersymmetric object.
Taking into account these heuristic arguments we define the residue of a
pseudo-differential operator ${\Psi}$ with respect to the fermionic
covariant derivative $D_+$ according to the rule:
\begin{eqnarray}
{\Psi} \equiv ...+(D_{-}res({\Psi}))D_{+}^{-1}+... 
\label{defres}
\end{eqnarray}
which will also be justified a posteriori. Then, bosonic and fermionic
Hamiltonian densities can be defined as:
\begin{eqnarray}
{\cal H}^{t}_l \equiv res(L^{-}_{2l}),
\label{res0}
\end{eqnarray}
\begin{eqnarray}
\quad \quad  ~{\cal H}^{U^{+}}_l \equiv res(N^{+}_{l}), \quad
{\cal H}^{U^{-}}_l \equiv res(N^{-}_{l} - {\widetilde N}^{-}_{2l+1}),
\label{resq1}
\end{eqnarray}
\begin{eqnarray}
{\cal H}^{+}_l \equiv
res(L^{+}_{2l-1}), \quad {\cal H}^{-}_l \equiv
res(L^{-}_{2l-1}-{\widetilde L}^{-}_{2l-1}),
\label{res}
\end{eqnarray}
\begin{eqnarray}
~~{\widetilde {\cal H}}^{+}_l \equiv res(M^{+}_{2l-1}), \quad
{\widetilde {\cal H}}^{-}_l \equiv res(M^{-}_{2l-1}-
{\widetilde M}^{-}_{2l-1}).
\label{res1}
\end{eqnarray}
Let us underline that all the operators generating the flows \p{satoflow} 
are included into these formulae.
Using these formulae and the relations (\ref{ar}--\ref{conj2}) 
one can derive the general formulae for the Hamiltonians 
${H}^{t}_l$, ${H}^{U^{-}}_l$, ${H}^{-}_l$ and
${\widetilde {H}}^{-}_l$ in terms of the Lax operator ${\cal L}$ \p{def1} 
modulo inessential factors\footnote{Let us recall that Hamiltonian densities
are defined up to terms which are fermionic or bosonic total derivatives
of an arbitrary functional $f(Z)$ of the initial superfields which,
however, should satisfy the following constraint:
$f(+\infty, \theta^+,\theta^-)-f(-\infty,\theta^+,\theta^-)=0$.}:
\begin{eqnarray}
&&{H}^{t}_{l-1} = 
\int dZ D_{-}^{-1}\sum_{k=0}^{2l-3}(-1)^{k}({\cal L}^{2l-3-k}v)
(({\cal L}^{k})^{T}u), \nonumber\\
&&{H}^{U^{-}}_{l-1} =  \int dZ D^{-1}_{-}\sum_{k=0}^{2l-3}
({\cal L}^{2l-3-k}v) (({\cal L}^{k})^{T}u), \nonumber\\
&&{H}^{-}_{l} =
\int dZ D_{-}^{-1}\sum_{k=0}^{2(l-1)}({\cal L}^{2(l-1)-k}v)
(({\cal L}^{k})^{T}u), \nonumber\\
&&{\widetilde H}^{-}_{l} = 
\int dZ D_{-}^{-1}\sum_{k=0}^{2(l-1)}(-1)^{k}({\cal L}^{2(l-1)-k}v)
(({\cal L}^{k})^{T}u).
\label{hamsgen}
\end{eqnarray}
We present, for example, the following explicit
expressions for the first few bosonic,
\begin{eqnarray}
&& \quad \quad \quad \quad \quad \quad \quad 
{H}^{t}_1 = \int dZ uv, \quad {H}^{t}_2 =\int dZ uv~', \nonumber\\
&&\quad \quad {H}^{t}_3 = \int 
dZ ~[~uv~''+vu[u(D_{+}D_{-}v)-v(D_{+}D_{-}u)]+\frac{2}{3}(uv)^3~],
\label{hams}
\end{eqnarray}
\begin{eqnarray}
&& \quad \quad {H}^{U^{\mp}}_1 = \int dZ ~[~\frac{1}{2}uv-{\theta}^{\mp}
(vD_{\mp}u \mp vuD_{\pm}^{-1}(uv))~],
\label{hamsq+-}
\end{eqnarray}
\begin{eqnarray}
&&\quad \quad {H}^{U}_1 = \int dZ ~[~{\theta}^{+}(vD_{-}u -
uvD_{+}^{-1}(uv))
+{\theta}^{-}(vD_{+}u + uvD_{-}^{-1}(uv))~], \nonumber\\
&& \quad \quad \quad \quad \quad {H}^{{\overline U}}_1 =\int dZ~
\frac{1}{2} ~u~[~{\theta}^{-}(D_{+}+Q_{+})-{\theta}^{+}(D_{-}+Q_{-})~]~v,
\label{hamsqbar}
\end{eqnarray}
and fermionic,
\begin{eqnarray}
\quad {H}^{\mp}_1= {\widetilde H}^{\mp}_1 = \int dZ D_{\mp}^{-1}(uv),
\label{hamsfermnonl}
\end{eqnarray}
\begin{eqnarray}
&& \quad \quad \quad
{H}^{\mp}_2 \equiv \frac{3}{2}H^{D^{\mp}}_2+\frac{1}{2}H^{Q^{\mp}}_2,
\quad {{\widetilde H}}^{\mp}_2\equiv
 \frac{1}{2}H^{D^{\mp}}_2+ \frac{3}{2}H^{Q^{\mp}}_2, \nonumber\\
&&\quad {H}^{D^{{\mp}}}_2= \int dZ ~[~\mp vD_{\mp}u +
uvD_{\pm}^{-1}(uv)~],
\quad {H}^{Q^{{\mp}}}_2 = \int dZ vQ_{\mp}u, \nonumber\\ 
&&{H}^{D^{{\mp}}}_3 = \int dZ ~[~ \mp vD_{\mp}u~' +
2vu~'D_{\pm}^{-1}(uv)+v(D_{\mp}u)D_{\pm}^{-1}D_{\mp}(uv)~],
\label{hamsferm}
\end{eqnarray}
Hamiltonians\footnote{When deriving eqs. (\ref{hamsgen}--\ref{hamsferm})
we integrated by parts and made essential use of realizations
\p{derrealiz} for the inverse derivatives and of the relationship 
$Q_{\pm}\equiv D_{\pm}-2{\theta}^{\pm}{\partial}$. We also used the
following definition of the superspace integral:
$\int dZ f(Z)\equiv \int dz (D_+D_-f)(z,0,0)$.}. The Hamiltonians
$H^{U}_1$ and $H^{\overline U}_1$ \p{hamsqbar} were found out by
hand so that they are conserved quantities with respect to the 
flows ${\textstyle{\partial\over\partial t_l}}$ \p{eqs}.
In eqs. \p{hamsferm} the Hamiltonians derived from
eqs. (\ref{res}--\ref{res1}) are presented
as a combination of the linearly independent Hamiltonians of
the basis \p{notat}.

Let us stress that the Hamiltonians in eqs. (\ref{res0}--\ref{res1}) 
are only conjectured to be conserved under the bosonic flows 
${\textstyle{\partial\over\partial t_l}}$ \p{lax}. This conjecture was
checked for a few of them under the explicit flows \p{eqs}.

It is well known that a bi-Hamiltonian system of
evolution equations can be represented as:
\begin{eqnarray}
{\textstyle{\partial\over\partial {\tau}^{a}_l}}
\left(\begin{array}{cc} v \\ u \end{array}\right) =
J_1 \left(\begin{array}{cc} {\delta}/{\delta v} \\
{\delta}/{\delta u} \end{array}\right) H^{a}_{l+1}=
J_2 \left(\begin{array}{cc} {\delta}/{\delta v} \\
{\delta}/{\delta u} \end{array}\right) H^{a}_{l},
\label{hameq}
\end{eqnarray}
where $J_1$ and $J_2$ are the first and second Hamiltonian structures.
In terms of these the Poisson brackets of the
superfields $v$ and $u$ are given by the formula:
\begin{eqnarray}
\{\left(\begin{array}{cc}
v(Z_1)\\ u(Z_1)\end{array}\right)
\stackrel{\otimes}{,}
\left(\begin{array}{cc} v(Z_2),u(Z_2)\end{array}\right)\}_l=
J_l(Z_1){\delta}^{N=2}(Z_1-Z_2).
\label{palg}
\end{eqnarray}
Using the flows (\ref{eqs}--\ref{qqq1})
and Hamiltonians (\ref{hams}--\ref{hamsferm}), we have found the first
\begin{eqnarray}
~~J_1= \left(\begin{array}{cc} 0 & 1 \\
-1 &  0\end{array}\right)
\label{hamstr}
\end{eqnarray}
and second
\begin{eqnarray}
&& \quad \quad \quad \quad \quad \quad \quad \quad \quad
J_2= \left(\begin{array}{cc} J_{11} & J_{12} \\
J_{21} &  J_{22} \end{array}\right), \nonumber\\
J_{11} &\equiv & +vD_{+}^{-1}vD_{-}-(D_{-}v)D_{+}^{-1}v-
2vD_{+}^{-1}uvD_{+}^{-1} v\nonumber\\ &&-vD_{-}^{-1}vD_{+}
+(D_{+}v)D_{-}^{-1}v-2vD_{-}^{-1}uvD_{-}^{-1} v,\nonumber\\
J_{22} & \equiv & +uD_{-}^{-1}uD_{+}-(D_{+}u)D_{-}^{-1}u-
2uD_{-}^{-1}uvD_{-}^{-1} u\nonumber\\ &&-uD_{+}^{-1}uD_{-}
+(D_{-}u)D_{+}^{-1}u-
2uD_{+}^{-1}uvD_{+}^{-1} u, \nonumber\\
J_{12} & \equiv & \partial + \{D_{-},vD_{+}^{-1}u\}+
2vD_{+}^{-1}uvD_{+}^{-1}u- \{D_{+},vD_{-}^{-1}u\}+
2vD_{-}^{-1}uvD_{-}^{-1}u,\nonumber\\
J_{21}& \equiv & \partial + \{D_{+},uD_{-}^{-1}v\}+
2uD_{-}^{-1}uvD_{-}^{-1}v- \{D_{-},uD_{+}^{-1}v\}+
2uD_{+}^{-1}uvD_{+}^{-1}v
\label{hamstr2}
\end{eqnarray}
Hamiltonian structures of the hierarchy.

The Jacobi identities for the first Hamiltonian structure $J_1$
\p{hamstr} are obviously satisfied.
The second Hamiltonian structure $J_2$ \p{hamstr2} is a very
complicated, nonlinear and nonlocal algebra, and due to this reason it is
a very nontrivial, technical task to verify its Jacobi identities.
It becomes a simpler linear algebra in terms of the original
Toda-lattice superfields $b$ and $f$ \p{zerocurveqs3}
\begin{eqnarray}
\quad \quad \quad \quad b\equiv uv, \quad f\equiv D_+\ln v.
\label{zerocurveqs3trans}
\end{eqnarray}
The corresponding
Hamiltonian structures $J^{(b,f)}_1$ and $J^{(b,f)}_2$ can be expressed
via $J_1$ and $J_2$ (\ref{hamstr}--\ref{hamstr2}) by the following
standard relation:
\begin{eqnarray}
\quad \quad \quad \quad
J^{(b,f)}_l= FJ_lF^{T}, \quad
F\equiv \left(\begin{array}{cc} u & v \\
D_{+}\frac{1}{v} &  0 \end{array}\right),
\label{hamstr-bfbasis}
\end{eqnarray}
where $F$ is the matrix of Frechet derivatives corresponding to the
transformation $\{b,f\} \Rightarrow \{v,u\}$  \p{zerocurveqs3trans}.
One finds:
\begin{eqnarray}
&& \quad \quad
J^{(b,f)}_1= \left(\begin{array}{cc} 0 & D_+ \\
D_{+} &  0 \end{array}\right), \quad
J^{(b,f)}_2= \left(\begin{array}{cc} J^{(b,f)}_{11} & J^{(b,f)}_{12} \\
J^{(b,f)}_{21} &  J^{(b,f)}_{22} \end{array}\right), \nonumber\\
J^{(b,f)}_{11} &\equiv & +{\partial} b + b {\partial}, \nonumber\\
J^{(b,f)}_{12} & \equiv & -{\partial}D_{+}+D_- b+D_{+}bD_+D_{-}^{-1}
+(D_+f)D_+, \nonumber\\
J^{(b,f)}_{21} & \equiv & +{\partial}D_{+}+bD_- -D_+D_{-}^{-1}bD_+
+D_+(D_+f), \nonumber\\
J^{(b,f)}_{22}& \equiv & -2D_{-}D_{+}+2b-2(D_-f)+[(D_+f),D_+D_{-}^{-1}]
-2D_{+}D_{-}^{-1}bD_{+}D_{-}^{-1}.
\label{hamstr2bf}
\end{eqnarray}
The Jacobi identities for $J^{(b,f)}_2$
\p{hamstr2bf} are still very complicated, and
we have only verified the consistency of the
Jacobi identities\footnote{The Jacobi identity $\{b,b,b\}$ is
satisfied because the restriction of the Poisson brackets to
the superfield $b$ forms the classical Virasoro algebra.}
$\{b,b,f\}$ and $\{b,f,f\}$. Taking into account that $J_2$ correctly
reproduces all the flows explicitly derived,
it is natural to expect that the most complicated, remaining Jacobi
identity $\{f,f,f\}$ is satisfied as well, but we cannot present a proof
here. One more argument in the favour of this expectation is given by
the secondary reduction of $J^{(b,f)}_2$ (see section 8) which coincides
with the $N=2$ superconformal algebra \p{redhamstr2bf}.

Using equations \p{hamsfermnonl}, \p{hameq} and \p{hamstr}, we obtain, for
example, the 0-th fermionic flow,
\begin{eqnarray}
\quad \quad D^{\pm}_0 v=v{\theta}^{\pm}, \quad
D^{\pm}_0 u=-u{\theta}^{\pm}.
\label{ff0} \end{eqnarray}

Knowledge of the first and second Hamiltonian structures allows us to
construct the recursion operator of the hierarchy using the following
general rule:
\begin{eqnarray}
R = J_2 J^{-1}_1 \equiv
\left(\begin{array}{cc}
J_{12}, & -J_{11} \\
J_{22}, &  -J_{21}
\end{array}\right), \quad
\frac{\partial}{\partial {\tau}^{a}_{l+1}}
\left(\begin{array}{cc} v\\u \end{array}\right) =
R \frac{\partial}{\partial {\tau}^{a}_{l}}
\left(\begin{array}{cc} v\\u \end{array}\right),
\label{recop0}
\end{eqnarray}
\begin{eqnarray}
{\textstyle{\partial\over\partial {\tau}^{a}_{l+1}}}
\left(\begin{array}{cc} v\\\ u \end{array}\right) =
R^{l} {\textstyle{\partial\over\partial {\tau}^{a}_{1}}}
\left(\begin{array}{cc} v\\u \end{array}\right), \quad
J_{l+1} = R^l J_1.
\label{hamstrn}
\end{eqnarray}
The Hamiltonian structures $J^{(b,f)}_1$ and $J^{(b,f)}_2$ \p{hamstr2bf}
(and, consequently the original Hamiltonian structures $J_1$ and $J_2$
(\ref{hamstr}--\ref{hamstr2}) ) are obviously compatible: the deformation
of the superfield $f$ $\Rightarrow$  $f + \gamma \theta^+$, where $\gamma$
is an arbitrary parameter, transforms $J^{(b,f)}_2$ into the Hamiltonian
structure 
\begin{eqnarray}
 \quad \quad \quad
J^{(b,f+\gamma\theta^+)}_2= J^{(b,f)}_2+\gamma J^{(b,f)}_1.
\label{hamstr-comp}
\end{eqnarray}
Thus, one concludes that the recursion operator $R$ \p{recop0} is
hereditary as the operator obtained from the compatible pair of 
Hamiltonian structures \cite{ff}.

Applying formulae \p{recop0} we obtain the following recurrence relations
for the flows:
\begin{eqnarray}
{\textstyle{\partial\over\partial {\tau}^{a}_{l+1}}}v=
+{\textstyle{\partial\over\partial {\tau}^{a}_{l}}}v~'
&+& (-1)^{d_{{\tau}^{a}}}vD_{+}^{-1}
{\textstyle{\partial\over\partial {\tau}^{a}_{l}}}
{\cal H}^{D^{{-}}}_2 +[(D_{-}v)+v(D_{+}^{-1}uv)]D_{+}^{-1}
{\textstyle{\partial\over\partial {\tau}^{a}_{l}}}{\cal H}^{t}_1 \nonumber\\
&+&(-1)^{d_{{\tau}^{a}}}vD_{-}^{-1}{\textstyle{\partial\over\partial
{\tau}^{a}_{l}}}
{\cal H}^{D^{{+}}}_2 -[(D_{+}v)-v(D_{-}^{-1}uv)]D_{-}^{-1}
{\textstyle{\partial\over\partial {\tau}^{a}_{l}}}{\cal H}^{t}_1, \nonumber\\
{\textstyle{\partial\over\partial {\tau}^{a}_{l+1}}}u=
-{\textstyle{\partial\over\partial {\tau}^{a}_{l}}}u~'
&-& (-1)^{d_{{\tau}^{a}}}uD_{-}^{-1}
{\textstyle{\partial\over\partial {\tau}^{a}_{l}}}
{\cal H}^{D^{{+}}}_2 -[(D_{+}u)+u(D_{-}^{-1}uv)]D_{-}^{-1}
{\textstyle{\partial\over\partial {\tau}^{a}_{l}}}{\cal H}^{t}_1 \nonumber\\
&-&(-1)^{d_{{\tau}^{a}}}uD_{+}^{-1}{\textstyle{\partial\over\partial
{\tau}^{a}_{l}}}{\cal H}^{D^{{-}}}_2 +
[(D_{-}u)-u(D_{+}^{-1}uv)]D_{+}^{-1}
{\textstyle{\partial\over\partial {\tau}^{a}_{l}}}{\cal H}^{t}_1 \nonumber\\
&-&2uD_{+}^{-1}D_{-}{\textstyle{\partial\over\partial
{\tau}^{a}_{l}}}{\cal H}^{t}_1,
\label{recrel}
\end{eqnarray}
where $d_{{\tau}^{a}}$ is the Grassmann parity of the evolution derivative
${\textstyle{\partial\over\partial {\tau}^{a}_{l}}}$ and 
\begin{eqnarray}
{\cal H}^{t}_1 \equiv uv, \quad
{\cal H}^{D^{{\mp}}}_2 \equiv \mp vD_{\mp}u + uvD_{\pm}^{-1}(uv)
\label{dens}
\end{eqnarray}
are the densities of the Hamiltonians ${H}^{t}_1$ \p{hams} and 
${H}^{D^{{\mp}}}_2$ \p{hamsferm}, respectively.

Taking into account the involution properties
\begin{eqnarray}
&& \quad \quad \quad \quad \quad \quad 
-{({\cal H}^{t}_1)}^{*}= {({\cal H}^{t}_1)}^{\dagger}=
{({\cal H}^{t}_1)}^{\star}= {({\cal H}^{t}_l)}^{\bullet}= 
{\cal H}^{t}_1, \nonumber\\
&&({\cal H}^{D^{{\mp}}}_2)^{*}=\pm D_{\mp}{\cal H}^{t}_1+
{\cal H}^{D^{{\mp}}}_2, \quad
({\cal H}^{D^{{\mp}}}_2)^{\dagger}= \mp D_{\pm}{\cal H}^{t}_1+
{\cal H}^{D^{{\pm}}}_2, \quad
({\cal H}^{D^{{\mp}}}_2)^{\star}= \pm {\cal H}^{D^{{\pm}}}_2
\label{conjf1}
\end{eqnarray}
of the Hamiltonian densities 
${\cal H}^{t}_1$ and ${\cal H}^{D^{{\mp}}}_2$ \p{dens},
one can verify that the recurrence relations (\ref{recrel}) possess
the involutions (\ref{conj*}--\ref{conbullet}). Together with the
fact already verified in section 5 that the first flows
(\ref{eqs}--\ref{qqq1})
also admit these involutions, one can conclude that all the flows of the
hierarchy under consideration admit them as well.
 
Using eqs. (\ref{recrel}), we obtain, for example, the 3rd bosonic flow
\begin{eqnarray}
{\textstyle{\partial\over\partial t_3}} v &=&
v~''' -3(D_{+}v)~'(D_{-}uv)+3(D_{-}v)~'(D_{+}uv)
-3v~'(D_{+}u)(D_{-}v) \nonumber\\ 
&&+3v~'(D_{-}u)(D_{+}v) 
-6vv~'(D_{+}D_{-}u)+6(uv)^2v~', \nonumber\\
{\textstyle{\partial\over\partial t_3}} u &=&
u~''' -3(D_{-}u)~'(D_{+}uv)+3(D_{+}u)~'(D_{-}uv)
-3u~'(D_{-}v)(D_{+}u) \nonumber\\ &&+3u~'(D_{+}v)(D_{-}u) 
-6uu~'(D_{-}D_{+}v)+6(uv)^2u~'
\label{flow3}
\end{eqnarray}
which coincides with the corresponding flow that can be derived from
the Lax-pair representation (\ref{lax}).

Let us end this section with the remark that due to the local nature of
bosonic flows \p{lax} and the relation
\begin{eqnarray} 
{\cal H}_1^{\mp} = D_{\mp}^{-1}{\cal H}^{t}_1
\label{rel1}
\end{eqnarray}
(see eqs. \p{hams} and \p{hamsfermnonl}), the evolution equations
for ${\cal H}^{t}_1$ with respect to the bosonic times $t_l$ should admit
the following very special, important representation:
\begin{eqnarray}
{\textstyle{\partial\over\partial t_l}}{\cal H}^{t}_1=h^{(1)}_l~'+
D_{+}D_{-}h^{(2)}_l,
\label{rel2}
\end{eqnarray}
where $h^{(1)}_l$ and $h^{(2)}_l$  are some local functions. We have
explicitly checked it for the first few values of $l$ and
observed that
\begin{eqnarray}
\quad \quad h^{(1)}_l = {\cal H}^{t}_l, \quad h^{(2)}_l = {\cal H}^{t}_1
{\cal H}^{t}_{l-1}
\label{rel3}
\end{eqnarray}
modulo an inessential factor and total derivatives. We also analyzed
a similar construction of fermionic flows \p{ff2-} and derived the
following formula:
\begin{eqnarray}
D^{\pm}_l {\cal H}^{t}_1= -{\cal H}^{\pm}_l~'.
\label{relf3}
\end{eqnarray}
These formulae are conjectured to be valid for any value of $l$,
consequently the
flows \p{recrel} allow the bosonic and fermionic Hamiltonian densities
${\cal H}^{t}_l$ \p{res0} and ${\cal H}^{\pm}_l$
\p{res} to be constructed using eqs.
(\ref{rel2}--\ref{relf3}), i.e. almost all information
about Hamiltonians and flows is encoded in the recursion operator
\p{recop0}.

\section{$U(1)$-symmetric basis}
In this section we introduce a new basis of the algebra of flows
(\ref{alg1}--\ref{algqqbar}), find $U(1)$--automorphism 
transformations for all reduced flows, and applying these
transformations derive a new Lax operator.

It is instructive to introduce the new basis
\begin{eqnarray}
\{ D_{\pm},Q_{\pm}, D^{\pm}_l,Q^{\pm}_l,u,v\} \Longrightarrow
\{D,{\overline D},Q,{\overline Q},D_l,{\overline D}_l,Q_l,
{\overline Q}_l,\frac{1}{\sqrt{i}}~u, \frac{1}{\sqrt{i}}~v\}, 
\label{N2basis}
\end{eqnarray}
\begin{eqnarray}
&& D\equiv \frac{1}{\sqrt {2}}(D_-+iD_+), \quad
{\overline D}\equiv \frac{1}{\sqrt {2}}(D_--iD_+), \nonumber\\
&&Q\equiv \frac{1}{\sqrt {2}}(Q_-+iQ_+), \quad
~{\overline Q}\equiv \frac{1}{\sqrt {2}}(Q_--iQ_+), \nonumber\\
&&D_k\equiv \frac{1}{\sqrt {2}}(D^{-}_k+i D^{+}_k), \quad
{\overline D}_k\equiv \frac{1}{\sqrt {2}}(D^{-}_k-i D^{+}_k),\nonumber\\ 
&&Q_k\equiv \frac{1}{\sqrt {2}}(Q^{-}_k+i Q^{+}_k), \quad
~{\overline Q}_k\equiv \frac{1}{\sqrt {2}}(Q^{-}_k-i Q^{+}_k)
\label{trans}
\end{eqnarray}
in the algebras (\ref{alg00}) and \p{alg1} which now become:
\begin{eqnarray}
\quad \quad \quad
\{D,{\overline D}\}=+2{\partial}, \quad \{Q,{\overline Q}\}=-2{\partial}, 
\label{N2der0}
\end{eqnarray}
\begin{eqnarray}
\quad \quad \quad
\Bigl\{D_k\,,\,{\overline D}_l\Bigr\}=
-2\;\frac{{\partial}}{{\partial t_{k+l-1}}}, \quad
\Bigl\{Q_k\,,\,{\overline Q}_l\Bigr\}=
+2\;\frac{{\partial}}{{\partial t_{k+l-1}}}, 
\label{N2alg0}
\end{eqnarray}
respectively, where $i$ is the imaginary unity.
Then, the first bosonic and fermionic flows from eqs.
(\ref{eqs}--\ref{supersflows}) and recurrence relations \p{recrel} become
\begin{eqnarray}
&&{\textstyle{\partial\over\partial t_1}}
\left(\begin{array}{cc} v\\ u \end{array}\right) =
{\partial}\left(\begin{array}{cc} v\\ u \end{array}\right), 
\quad Q_1 \left(\begin{array}{cc} v\\ u \end{array}\right) =
Q\left(\begin{array}{cc} v\\ u \end{array}\right), \quad
{\overline Q}_1 \left(\begin{array}{cc} v\\ u \end{array}\right) =
{\overline Q}\left(\begin{array}{cc} v\\ u \end{array}\right), \nonumber\\
&& \quad \quad \quad D_1 v=-Dv-2v{\partial}^{-1}D(uv),\quad
D_1 u=-Du+2u{\partial}^{-1}D(uv),\nonumber\\
&&\quad \quad \quad 
{\overline D}_1 v=-{\overline D}v+2v{\partial}^{-1}{\overline D}(uv),\quad
{\overline D}_1 u=-{\overline D}u-2u{\partial}^{-1}{\overline D}(uv), 
\label{N2supersflows}
\end{eqnarray}
\begin{eqnarray}
{\textstyle{\partial\over\partial {\tau}^{a}_{l+1}}}v&=&
+{\textstyle{\partial\over\partial {\tau}^{a}_{l}}}v~' \nonumber\\
&+&(-1)^{d_{{\tau}^{a}}} [v{\partial}^{-1}D
{\textstyle{\partial\over\partial {\tau}^{a}_{l}}}
(v{\overline D}u-uv{\partial}^{-1}{\overline D}uv)
-v{\partial}^{-1}{\overline D}
{\textstyle{\partial\over\partial {\tau}^{a}_{l}}}
(vDu+uv{\partial}^{-1}Duv)] \nonumber\\
&-&[({\overline D}v)+v({\partial}^{-1}{\overline D}uv)]{\partial}^{-1}D
{\textstyle{\partial\over\partial {\tau}^{a}_{l}}}(uv)
+[(Dv)-v({\partial}^{-1}Duv)]{\partial}^{-1}{\overline D}
{\textstyle{\partial\over\partial {\tau}^{a}_{l}}}(uv),\nonumber\\
{\textstyle{\partial\over\partial {\tau}^{a}_{l+1}}}u&=&
-{\textstyle{\partial\over\partial {\tau}^{a}_{l}}}u~'\nonumber\\
&-&(-1)^{d_{{\tau}^{a}}}[u{\partial}^{-1}
{\overline D}{\textstyle{\partial\over\partial {\tau}^{a}_{l}}}
(uDv-uv{\partial}^{-1}Duv)-
u{\partial}^{-1}D{\textstyle{\partial\over\partial {\tau}^{a}_{l}}}
(u{\overline D}v+uv{\partial}^{-1}{\overline D}uv)] \nonumber\\
&+&[(Du)+u({\partial}^{-1}Duv)]{\partial}^{-1}{\overline D}
{\textstyle{\partial\over\partial {\tau}^{a}_{l}}}(uv)
-[({\overline D}u)-u({\partial}^{-1}{\overline D}uv)]{\partial}^{-1}D
{\textstyle{\partial\over\partial {\tau}^{a}_{l}}}(uv), 
\label{recrelN2basis}
\end{eqnarray}
respectively, and their simple inspection shows that they admit the 
$U(1)$ automorphism of the $N=2$ supersymmetry algebras
(\ref{N2der0}--\ref{N2alg0}) hidden in the former basis,
\begin{eqnarray}
&& (\frac{{\partial}}{\partial t_l})
\quad \quad \quad \quad \quad \quad \quad \quad \quad \Longrightarrow \quad
\quad \quad \quad \quad
(\frac{{\partial}}{\partial t_l}), \nonumber\\
&&(D,\quad Q, \quad D_l, \quad Q_l)\quad \Longrightarrow
\quad \exp{(+i\phi)}~ (D, \quad Q, \quad D_l, \quad Q_l), \nonumber\\
&&({\overline D},\quad {\overline Q},\quad {\overline D}_l,\quad
{\overline Q}_l) \quad
\Longrightarrow \quad \exp{(-i\phi)}~({\overline D},\quad {\overline Q},
\quad {\overline D}_l,\quad {\overline Q}_l),
\label{N2u1}
\end{eqnarray}
where $\phi$ is an arbitrary parameter. Consequently, all higher
flows admit this automorphism as well.  Without going into additional
technical details let us also present the corresponding
$U(1$)--transformations for the remaining bosonic flows of the hierarchy
\begin{eqnarray}
&& (U^{+}_l+U^{-}_l, \quad {\overline U}_l) \quad \Longrightarrow \quad
\quad \quad \quad \quad \quad
(U^{+}_l+U^{-}_l, \quad {\overline U}_l),\nonumber\\
&&(U^{+}_l-U^{-}_l\pm iU_l) \quad \Longrightarrow \quad \exp{(\mp 2i\phi)}~
(U^{+}_l-U^{-}_l \pm i U_l).
\label{N2qu1}
\end{eqnarray}
As we already mentioned at the end of section 4 the Lax operator 
${\cal L}$ \p{def1} is not invariant with respect to 
$U(1)$ transformations, therefore applying these to ${\cal L}$
one can derive the one-parameter family of consistent Lax operators
\begin{eqnarray}
\quad \quad \quad \quad
{\cal L} \quad \Longrightarrow \quad {\cal L}^{\phi}=
\cos \phi ~{\cal L}+ \sin \phi ~{\cal L}^{\star}
\label{ulax}
\end{eqnarray}
which generate isomorphic flows, where ${\cal L}^{\star}$ is defined
in eq. \p{conjlax}.

For completeness, we present also the explicit expressions of the
first and second Hamiltonian structures in the special superfield basis
\begin{eqnarray}
\quad \quad \quad \quad \{u, \quad v\}  \quad \Longrightarrow \quad \{b
\equiv iuv, \quad {\widetilde b}\equiv (\ln v)~'\},
\label{N2basissup}
\end{eqnarray}
where they possess both a manifest $N=2$ supersymmetry and form
linear superalgebras,
\begin{eqnarray}
&&  \quad \quad \quad \quad
J^{(b,{\widetilde b})}_1= \left(\begin{array}{cc} 0 & {\partial} \\
{\partial} &  0 \end{array}\right), \quad
J^{(b,{\widetilde b})}_2= \left(\begin{array}{cc} 
J^{(b,{\widetilde b})}_{11} &
J^{(b,{\widetilde b})}_{12} \\
J^{(b,{\widetilde b})}_{21} &  J^{(b,{\widetilde b})}_{22}
\end{array}\right), \nonumber\\
J^{(b,{\widetilde b})}_{11} &\equiv & + {\partial}b+b{\partial},
\nonumber\\ J^{(b,{\widetilde b})}_{12} & \equiv & -{\partial}^2-
{\overline D}bD + Db{\overline D} + {\widetilde b}{\partial}, \nonumber\\
J^{(b,{\widetilde b})}_{21} & \equiv & + {\partial}^2+
{\overline D}bD - Db{\overline D} + {\partial}{\widetilde b},\nonumber\\
J^{(b,{\widetilde b})}_{22}& \equiv &
+[D,{\overline D}~]~{\partial} -2 Db{\overline D} -2 {\overline D}bD +
(D~{\widetilde b}){\overline D} -({\overline D}~{\widetilde b})D -
([D,{\overline D}~]{\partial}^{-1}{\widetilde b})~{\partial}.
\label{hamstr2bfN2}
\end{eqnarray}
The last superalgebra is minimally nonlocal: it contains only a single
term with an inverse derivative in the  $J^{(b,{\widetilde b})}_{22}$
component. It seems that there is no superfield basis where it is possible
to avoid this nonlocality. It would be interesting to clarify its origin
in the framework of Hamiltonian reduction of
affine superalgebras, but this question is still unclear now.

\section{Secondary reduction: the $N=2$ supersymmetric \\
${\alpha}=-2$ KdV hierarchy} 
In this section, by means of a secondary reduction, 
we establish a relationship between the $N=4$ Toda chain  
and $N=2$, ${\alpha}=-2$ KdV hierarchies.

Let us study the secondary reduction of the $N=4$ Toda chain hierarchy
considered in the preceding sections. With this aim, we impose the
following secondary constraint\footnote{See also
refs. \cite{yu,rs,ra}, where the similar reduction of the Manin-Radul
\cite{mr} and Mulase-Rabin \cite{m,r} $N=1$ supersymmetric KP and KdV
hierarchies has been discussed.} on the Lax operator ${\cal L}$ \p{def1}:
\begin{eqnarray}
{\cal L}^{T} = D_{+}{\cal L}D_{+}^{-1}
\label{red}
\end{eqnarray}
which can easily be resolved in terms of the superfield $v$ 
entering ${\cal L}$,
\begin{eqnarray}
v=1.
\label{red1}
\end{eqnarray}
Then, the reduced Lax operator ${\cal L}^{red}$ becomes
\begin{eqnarray}
{\cal L}^{red}= D_- +  D_{+}^{-1} u.
\label{red3}
\end{eqnarray}
The condition \p{red} by means of eq. \p{def1eq} induces the secondary
constraint
\begin{eqnarray}
({{\cal W}^{-1}})^{T} = D_{+}{\cal W}D_{+}^{-1}
\label{red4}
\end{eqnarray}
on the dressing operator ${\cal W}$ (\ref{def11}) 
which in turn induces the following secondary constraints on the operators
$L^{\pm}_l$, $M^{\pm}_l$,  $N^{\pm}_l$ \p{laxkp}:
\begin{eqnarray}
&& \quad \quad \quad \quad \quad \quad \quad \quad \quad 
(L^{\pm}_{2l})^{T} = (-1)^{l} D_{+}L^{\pm}_{2l}D_{+}^{-1}, \nonumber\\ 
&& (L^{\pm}_{2l-1})^{T} = (-1)^{l-\frac{1}{2} \pm \frac{1}{2}}
D_{+}L^{\pm}_{2l-1}D_{+}^{-1}, \quad
(M^{\pm}_{2l-1})^{T} = (-1)^{l-1}D_{+}M^{\pm}_{2l-1}D_{+}^{-1},\nonumber\\
&&(N^{-}_{l})^{T} = (-1)^{l+1} D_{+}(N^{-}_{l}-L^{-}_{2l})D_{+}^{-1},
\quad (N^{+}_{l})^{T} = (-1)^{l}  D_{+} N^{+}_{l}D_{+}^{-1} ~~~
\label{red5}
\end{eqnarray}
which are identically satisfied if constraint \p{red} (or \p{red4})
is imposed. The following important consequences obviously
result from eqs. \p{red5}:
\begin{eqnarray}
(L^{\pm}_{2(2k-1)})_{0} = (L^{+}_{2(2k)-1})_{0} = (M^{+}_{2(2k-1)-1})_{0}=
(N^{+}_{2k-1})_{0}=0, \quad k= 1,2 \ldots ,
\label{red6}
\end{eqnarray}
where the subscript $0$ refers to the constant part of the operators.
Consequently, the following equations:
\begin{eqnarray}
((L^{\pm}_{2(2k-1)})_{+}1) = ((L^{+}_{2(2k)-1})_{+}1) =
((M^{+}_{2(2k-1)-1})_{+}1)=((N^{+}_{2k-1})_{+}1)=0 
\label{red7}
\end{eqnarray}
are identically satisfied as well. Using these relations, the involution
\p{conjstar} and the algebra structure
(\ref{alg1}--\ref{algqqbar}) we are led to the conclusion that only half
of the flows (\ref{modflv}) are consistent with the reduction
(\ref{red}--\ref{red1}), and these flows are
\begin{eqnarray}
\{~{\textstyle{\partial\over\partial t_{2k-1}}},
~U^{\mp}_{2k-1},~U_{2k-1},~{\overline U}_{2k},~
D^{\mp}_{2k},~ Q^{\mp}_{2k-1}~\}.
\label{consistfl}
\end{eqnarray}

In order to understand deeper what kind of reduced hierarchy we have
in fact derived, let us analyze its Hamiltonian structure via Hamiltonian
reduction of the first and second Hamiltonian structures \p{hamstr} and
\p{hamstr2}, respectively, of the original hierarchy we started with.
It is easier to reduce the less complicated expressions
\p{hamstr2bf} in terms of the superfields $b,f$ \p{zerocurveqs3trans}. 
In this  basis constraint \p{red1} becomes 
\begin{eqnarray} 
f=0,
\label{red2} 
\end{eqnarray}
and on the constraint shell the superfield $b$ coincides with
the superfield $u$.

Let us start with the first Hamiltonian structure $J^{(b,f)}_{1}$
\p{hamstr2bf}. In this case constraint \p{red2} is a gauge
constraint, and the gauge can be fixed by the condition $b=0$. Thus, as a
result a trivial reduced Hamiltonian structure is generated.

In the case of the second Hamiltonian structure
$J^{(b,f)}_{2}$ \p{hamstr2bf}, constraint \p{red2} is second
class, and we can use the Dirac brackets in order to obtain
the second Hamiltonian structure of the reduced system. The result is
\begin{eqnarray}
{J^{(Dirac)}_{11}}=J^{(u,0)}_{11}-
J^{(u,0)}_{12}{J^{(u,0)}_{22}}^{-1}J^{(u,0)}_{21}\equiv
\frac{1}{2}({\partial}D_{+}D_{-}-D_-uD_{-}-D_{+}uD_{+}
+2{\partial} u + 2u {\partial}), ~ ~ ~
\label{redhamstr2bf}
\end{eqnarray}
where the relations
\begin{eqnarray}
J^{(b,0)}_{12}D_{-}=\frac{1}{2}
D_{-}J^{(b,0)}_{22}D_{-}=D_{-}J^{(b,0)}_{21}= -{\partial}D_{+}D_{-}+D_-
bD_{-}+D_{+}bD_{+}, 
\label{red1hamstr2bf}
\end{eqnarray}
which can easily be read from eqs. \p{hamstr2bf}, have been exploited. 

   From eq. \p{redhamstr2bf} we see that the second Hamiltonian structure
of the reduced hierarchy coincides with the $N=2$ superconformal algebra,
and from this remarkable fact one can conclude that
the reduced hierarchy should reproduce one of the three existing $N=2$
supersymmetric KdV hierarchies \cite{lm}. In order to establish which, 
let us derive the third flow
${\textstyle{\partial\over\partial t_3}}$ of the reduced hierarchy
by substituting constraint \p{red1} into equations \p{flow3}.
Then they become
\begin{eqnarray}
{\textstyle{\partial\over\partial t_3}} u &=&
(u~'' +3(D_{+}u)(D_{-}u)+2u^3)~'.
\label{redflow3}
\end{eqnarray}
Now, one can easily recognize that this is the third flow of the
so called $N=2$, ${\alpha}=-2$ KdV hierarchy \cite{lm}.
Thus, the reduced hierarchy is the $N=2$ supersymmetric
${\alpha}=-2$ KdV hierarchy.

To close this section we would like to
recall that different Lax--pair representations of the bosonic flows
${\textstyle{\partial\over\partial t_{2l-1}}}$ of
the $N=2$ supersymmetric ${\alpha}=-2$ KdV hierarchy were
discussed in \cite{lm,pop, dg1, bks2}, its recursion operator has been
constructed in \cite{op}, and the nonlocal fermionic flows were analyzed in
\cite{dmat}, while their Lax-pair description was missing. Here, as a
byproduct we derived the new Lax operator ${\cal L}^{red}$ \p{red3} for
this hierarchy which leads to a consistent Lax-pair representation of both
the bosonic and fermionic flows, and obtained the extended series of the
hierarchy of flows \p{consistfl}.

\section{Generalizations}
The hierarchies discussed in preceding sections admit a natural
generalization on the non-abelian case. Thus, we propose supersymmetric
matrix KP hierarchy generated by the matrix dressing operator $W$
\begin{eqnarray}
W \equiv I+\sum_{n=1}^\infty ~(w^{(0)}_n + 
w^{(+)}_nD_{+}+w^{(-)}_nD_{-}+ w^{(1)}_n D_{+}D_{-})~{\partial}^{-n},
\label{matrixkp}
\end{eqnarray}
and its consistent reductions characterized by the reduced operator
$L^{-}_1$
\begin{eqnarray}
L^{-}_{1}= ID_- + vD_{+}^{-1}u.
\label{matrixred}
\end{eqnarray}
Here, $w_n\equiv (w_n)_{AB}(Z)$, $v\equiv v_{Aa}(Z)$ and $u\equiv
u_{aA}(Z)$
($A,B=1,\ldots, k$; $a,b=1,\ldots , n+m$) are
rectangular matrix-valued superfields,
respectively, and $I$ is the unity matrix, $I\equiv {\delta}_{A,B}$.
In \p{matrixred} a matrix product is understood, for
example $(vu)_{AB} \equiv \sum_{a=1}^{n+m} v_{Aa}u_{aB}$.
The matrix entries are bosonic superfields for $a=1,\ldots ,n$ and
fermionic superfields for $a=n+1,\ldots , n+m$, i.e.,
$v_{Aa}u_{bB}=(-1)^{d_{a}{\overline d}_{b}}u_{bB}v_{Aa}$, where $d_{a}$
and ${\overline d}_{b}$ are the Grassmann
parities of the matrix elements $v_{Aa}$ and $u_{bB}$,
respectively, $d_{a}=1$ $(d_{a}=0)$ for fermionic (bosonic) entries.
The grading choosen guarantees that the Lax operator
$L_{1}^{-}$ is Grassmann odd \cite{bks2}.

The detailed analysis of the corresponding hierarchies is however out of 
the scope of the present paper and will be discussed elsewhere.
Nevertheless, let us only present a few first nontrivial bosonic and
fermionic flows in the noncommutative, matrix case (compare with the
abelian flows  (\ref{eqs}--\ref{ff2-}) )
\begin{eqnarray}
{\textstyle{\partial\over\partial t_2}} v =
+v~'' +  2\{D_{-},vD_{+}u\}v+2v(uv)^2, & \quad &
{\textstyle{\partial\over\partial t_2}} u =
-u~'' -  2\{D_{+},uD_{-}v\}u-2(uv)^2u, \nonumber\\
D^{+}_1 v= -D_+v+2(D^{-1}_-v{\cal I}u)v, & \quad &
D^{+}_1 u=-D_+u-2uD^{-1}_-(vu), \nonumber\\
D^{-}_1 v= -D_-v-2vD^{-1}_+(uv), & \quad & 
D^{-}_1 u= -D_-u+2{\cal I} (D^{-1}_+uv)u
\label{eqsmatrix}
\end{eqnarray}
which are derived using Lax-pair representations (\ref{lax}--\ref{modfl2})
with $L^{-}_1$ \p{matrixred} and
\begin{eqnarray}
(L^{+}_1)_{+} = ID_{+}-2(D_{-}^{-1}(v{\cal I}u)),
\label{Mmatrix}
\end{eqnarray}
and the matrix ${\cal I}$ is defined as
\begin{eqnarray}
{\cal I} \equiv (-1)^{d_a} {\delta}_{ab}.
\label{matrI}
\end{eqnarray}

To close this section let us only remark that for the particular case 
when the index $A=1$
the matrix reduced Lax operator \p{matrixred} becomes the scalar
operator generating the reduced hierarchy with the $n+m$ pairs
of the scalar superfields $v_a,u_a$. In the very particular case when the
indices $A=1,a=1$ and $n=1, m=0$ the Lax operator \p{matrixred} reproduces
the Lax operator \p{def1} of the $N=4$ supersymmetric Toda chain
hierarchy.

\section{Conclusion}
In this paper we have defined the $N=4$ generalization of
the $N=2$ supersymmetric KP hierarchy and derived its flows
and their algebra in the framework of the dressing approach. 
Then we have analyzed the possibility of viewing the
supersymmetric Toda chain hierarchy as a reduction of the $N=4$ KP
hierarchy and restored all $N=4$ KP reduced flows including the flows
of the $N=4$ supersymmetry. Due to this it is called 
the $N=4$ supersymmetric Toda chain hierarchy. Furthermore we have
exhibited its finite and infinite discrete symmetries and using them
derived its solutions and new Lax operators
generating isomorphic flows. Then we have explicitly calculated its first 
two Hamiltonian structures and recursion operator connecting all
its systems of evolution equations and Hamiltonian structures. Then we
have established its secondary reduction to the $N=2$ supersymmetric
${\alpha}=-2$ KdV hierarchy. Finally we have proposed a matrix
generalization of the $N=4$ supersymmetric KP hierarchy and
described an infinite family of its reductions characterized by
a finite number of superfields.

{}~

{}~

\noindent{\bf Acknowledgments.}
L.G. and A.S. thanks the Laboratoire de Physique Th\'eorique -
de l'ENS Lyon for the hospitality during the course of this work.
This work was partially supported by the PICS Project No. 593, 
RFBR-CNRS Grant No. 98-02-22034, RFBR Grant No. 99-02-18417,
Nato Grant No. PST.CLG 974874 and the Programme Emergence de la region
Rh\^one-Alpes (France).

\newpage

\section*{\bf Appendix}
\setcounter{equation}{0}
\def\theequation{A.\arabic{equation}}

Algebra (\ref{alg1}--\ref{algqqbar})
may be realized in the superspace $\{t_k,\theta^{\pm}_k,\rho^{\pm}_k,
h^{\pm}_k,h_k, {\overline h}_k \}$,
\begin{eqnarray}
&&D^{\pm}_k=
\frac{\partial}{\partial \theta^{\pm}_k}-
\sum^{\infty}_{l=1}\theta^{\pm}_l
\frac{\partial}{{\partial t_{k+l-1}}},\quad
Q^{\pm}_k=
\frac{\partial}{\partial {\rho}^{\pm}_k}+
\sum^{\infty}_{l=1}{\rho}^{\pm}_l
\frac{\partial}{{\partial t_{k+l-1}}}, 
\label{covder}
\end{eqnarray}
\begin{eqnarray}
U^{\pm}_k & = &
\frac{\partial}{\partial h^{\pm}_k}-
\sum^{\infty}_{l=1}({\theta}^{\pm}_l
\frac{\partial}{{\partial {\rho}^{\pm}_{k+l}}}+{\rho}^{\pm}_l
\frac{\partial}{{\partial {\theta}^{\pm}_{k+l}}}), \nonumber\\
U_k &=& \sum^{\infty}_{l=1}
\Bigl \{i({\overline u}_{l-1}
\frac{\partial}{\partial {\overline h}_{k+l-1}}-
u_{l-1}\frac{\partial}{\partial h_{k+l-1}})+\frac{1}{2}
(\frac{\partial}{\partial {\overline h}_{0}}{\overline u}_{l-1}+
\frac{\partial}{\partial h_{0}}u_{l-1})
(\frac{\partial}{\partial h^{+}_{k+l-1}}-\frac{\partial}{\partial
h^{-}_{k+l-1}}) \nonumber\\
&+& {\theta}^{+}_l\frac{\partial}{{\partial {\rho}^{-}_{k+l}}}
+{\theta}^{-}_l\frac{\partial}{{\partial {\rho}^{+}_{k+l}}}
+{\rho}^{+}_l\frac{\partial}{{\partial {\theta}^{-}_{k+l}}}
+{\rho}^{-}_l\frac{\partial}{{\partial {\theta}^{+}_{k+l}}}\Bigl \},
\nonumber\\
{\overline U}_k&=& -\sum^{\infty}_{l=1}
\{{\overline u}_{l-1}\frac{\partial}{\partial {\overline h}_{k+l-1}}+
u_{l-1}\frac{\partial}{\partial h_{k+l-1}}-\frac{i}{2}
(\frac{\partial}{\partial {\overline h}_{0}}{\overline u}_{l-1}-
\frac{\partial}{\partial h_{0}}u_{l-1})
(\frac{\partial}{\partial h^{+}_{k+l-1}}-\frac{\partial}{\partial
h^{-}_{k+l-1}})\nonumber\\
&-& {\theta}^{+}_l\frac{\partial}{{\partial {\theta}^{-}_{k+l}}}
+{\theta}^{-}_l\frac{\partial}{{\partial {\theta}^{+}_{k+l}}}
-{\rho}^{+}_l\frac{\partial}{{\partial {\rho}^{-}_{k+l}}}
+{\rho}^{-}_l\frac{\partial}{{\partial {\rho}^{+}_{k+l}}}\Bigl \},
\label{covder1}
\end{eqnarray}
where $t_k, h^{\pm}_k,h_k,{\overline h}_k$
($\theta^{\pm}_k,\rho^{\pm}_k$) are bosonic (fermionic) abelian
evolution times with dimensions
\begin{eqnarray}
[t_k]=[h^{\pm}_k]=[h_k]=[{\overline h}_k]=k, \quad
[\theta^{\pm}_k] =[\rho^{\pm}_k]=k-\frac{1}{2}.
\label{dim}
\end{eqnarray}
In eqs. \p{covder1} we have introduced the functions
\begin{eqnarray}
&&{\overline u}_l\equiv \frac{1}{l!}
\frac{\partial^{l}}{\partial {\sigma}^{l}}\Bigl \{ (1+
\sum^{\infty}_{k=0}\sum^{\infty}_{n=0}{\sigma}^{k+n}h_{k}{\overline h}_{n})
\exp{ [+i \sum^{\infty}_{m=0}{\sigma}^{m}(h^{+}_{m}-h^{-}_{m})]}\Bigl
\}\Bigl |_{{\sigma}=0}, \nonumber\\
&&u_l\equiv \frac{1}{l!}
\frac{\partial^{l}}{\partial {\sigma}^{l}}\Bigl \{ (1+
\sum^{\infty}_{k=0}\sum^{\infty}_{n=0}{\sigma}^{k+n}h_{k}{\overline h}_{n})
\exp{ [-i \sum^{\infty}_{m=0}{\sigma}^{m}(h^{+}_{m}-h^{-}_{m})]}\Bigl
\}\Bigl |_{{\sigma}=0}, \quad l=0,1,\ldots \nonumber\\
&& \quad \quad \quad \quad \quad \quad \quad \quad
{\overline u}_l\equiv 0, \quad u_l\equiv 0, \quad l< 0
\label{definition}
\end{eqnarray}
possessing the folowing properties:
\begin{eqnarray}
\frac{\partial}{\partial {x}_{k}}{\overline u}_{l}\equiv
\frac{\partial}{\partial {x}_{0}}{\overline u}_{l-k}, \quad
\frac{\partial}{\partial {x}_{k}}u_{l}\equiv
\frac{\partial}{\partial {x}_{0}}u_{l-k}, \quad
\frac{\partial}{\partial {h}^{\pm}_{0}}{\overline u}_{l}\equiv
\pm i{\overline u}_{l}, \quad
\frac{\partial}{\partial {h}^{\pm}_{0}}u_{l}\equiv \mp iu_{l}
\label{propert}
\end{eqnarray}
which together with the formula
\begin{eqnarray}
\frac{{\partial}^{l}}{\partial {\sigma}^{l}}(ab)\equiv
\sum^{l}_{k=0} \frac{l!}{(l-k)!k!}
(\frac{{\partial}^{k}}{\partial {\sigma}^{k}}a)
(\frac{{\partial}^{l-k}}{\partial {\sigma}^{l-k}}b)
\label{propert1}
\end{eqnarray}
can be used to check that the generators
(\ref{covder}--\ref{covder1})
indeed satisfy the algebra (\ref{alg1}--\ref{algqqbar}).
Here, $x_k$ is any evolution time from
the set $\{h^{\pm}_k,h_k,{\overline h}_k\}$ and $a,b$ are arbitrary
functions of some argument $\sigma$

{}~


\end{document}